\documentclass[preprint,12pt]{elsarticle}

\usepackage{amssymb}
\usepackage{amsthm}
\usepackage{amstext}
\usepackage{amsmath}
\usepackage{amsfonts}
\usepackage{breqn}
\setkeys{breqn}{breakdepth={0}}
\usepackage{booktabs}
\usepackage{bbm}
\usepackage{comment}
\usepackage[german,english]{babel}
\usepackage{epstopdf}
\usepackage{multirow}
\usepackage{tikz}
\usepackage{verbatim}

\usepackage{listings}

\newcommand\version{{v1.1.0.1}}
\newcommand\versionexample{{v1.1.0.1}}
\newcommand\C{\mathcal{C}}
\newcommand\cut{\text{-cut}}
\newcommand\NeatIBP{{\sc NeatIBP}}
\newcommand\FireFly{{\sc FireFly}}
\newcommand\Kira{{\sc Kira}}
\newcommand\Singular{{\sc Singular}}
\newcommand\Mathematica{{\sc Mathematica}}
\newcommand\dif{{\rm d}}
\newcommand\mc{{\text{mc}}}
\newcommand\Gr{{Gr{\"o}bner}}
\newcommand\preprint[1]{\gdef\@preprint{\hfill #1}}
\newcommand\tn\textnormal
\newcommand\spcref{\cite{Britto:2004nc,Larsen:2015ped,Ita:2015tya,Boehm:2020ijp,Georgoudis:2016wff,Bern:1994cg,Berger:2008sj,Bern:1997sc,Bern:2004cz,Bern:2000dn,Bern:1994zx,Zhang:2016kfo}}

\biboptions{sort&compress}

\newcounter{bla}

 \lstdefinestyle{yaml}{
  basicstyle=\ttfamily\small,
  commentstyle=\color{gray},
  stringstyle=\color{purple},
  keywordstyle=\color{magenta},
  morecomment=[l]{\#},
  morestring=[b]',
  morestring=[b]",
  morekeywords={true,false,null,yes,no,on,off},
  literate=%
    {---}{{\textcolor{red}{---}}}3
    {|}{{\textcolor{blue}{|}}}1
    {>}{{\textcolor{blue}{>}}}1
    {\&}{{\textcolor{magenta}{\&}}}1
    {*}{{\textcolor{magenta}{*}}}1
    { !!}{{\textcolor{blue}{ !!}}}3
    { :}{{\textcolor{blue}{ :}}}2
    { -}{{\textcolor{blue}{ -}}}2
}

\journal{Computer Physics Communications}

\usepackage{enumitem}
\usepackage[linesnumbered,ruled,noline]{algorithm2e}

\usepackage{tikz}
\usetikzlibrary{positioning}
\usetikzlibrary{arrows.meta}
\usetikzlibrary{shapes}
\usepackage{hyperref}
\usepackage{listings}
\usepackage{xcolor}
\usepackage{todonotes}
\definecolor{codegreen}{rgb}{0,0.6,0}
\definecolor{codegray}{rgb}{0.5,0.5,0.5}
\definecolor{codepurple}{rgb}{0.58,0,0.82}
\definecolor{backcolour}{rgb}{0.93,0.93,0.93}

\theoremstyle{plain}

\theoremstyle{definition}

\definecolor{ultramarine}{rgb}{0.0,0.28,0.68}

\usetikzlibrary{shapes,arrows}
\usetikzlibrary{decorations.pathmorphing}
\tikzstyle{startstop} = [rectangle,rounded corners=0.5cm, minimum width=2.5cm,minimum height=1cm,text centered, draw=black]
\tikzstyle{process} = [rectangle,rounded corners=0.2cm,minimum width=3cm,minimum height=1cm,text centered,text width=3cm,draw=black]
\tikzstyle{io} = [rectangle,minimum width=3cm,minimum height=1cm,text centered,text width =3cm,draw=black]
\tikzstyle{plaintext} = [rectangle,text centered,text width=3cm,minimum width=1cm,minimum height=0.5cm]
\tikzstyle{decision} = [trapezium, trapezium left angle = 70,trapezium right angle=110,minimum width=1cm,minimum height=0.5cm,text centered,text width=3cm,draw=black]
\tikzstyle{item} = [ellipse,minimum width=0.5cm,minimum height=1cm,text centered,text width=1.5cm,draw=black]
\tikzstyle{item2} = [ellipse,minimum width=0.5cm,minimum height=1cm,text centered,text width=0.5cm,draw=black]
\tikzstyle{arrow} = [thick,->,>=stealth]
\tikzstyle{arrow2} = [thick,<->,>=stealth]

\hypersetup{
    colorlinks=true,
    linkcolor=blue,
    citecolor=blue,
    urlcolor=blue
}

\SetKw{KwTrue}{True}
\SetKw{KwFalse}{False}
\SetKw{KwBreak}{break}
\SetKw{KwContinue}{continue}

\begin{document} 

\begin{frontmatter}

\title{ Performing integration-by-parts reductions using {\sc NeatIBP} 1.1 + {\sc Kira}}
\tnotetext[tnote1]{Report number: USTC-ICTS/PCFT-25-10, MPP-2025-29, HU-EP-25/12-RTG}

\author[HIAS]{Zihao Wu}
\ead{wuzihao@mail.ustc.edu.cn}
\author[RPTU]{Janko B{\"o}hm }
\ead{boehm@mathematik.uni-kl.de}
\author[PCFT,ICTS,MPI]{Rourou Ma}
\ead{marr21@mail.ustc.edu.cn}
\author[HU-IRIS]{Johann Usovitsch}
\ead{jusovitsch@gmail.com}
\author[HU]{Yingxuan Xu}
\ead{yingxu@physik.hu-berlin.de}
\author[PCFT,ICTS,PKU]{Yang Zhang}
\ead{yzhphy@ustc.edu.cn}

\address[HIAS]{Hangzhou Institute of Advanced study, University of Chinese Academy of Science, Hangzhou, Zhejiang 310000, China}
\address[RPTU]{Department of Mathematics, Rheinland-Pf\"alzische Technische Universit\"at Kaiserslautern-Landau (RPTU), 67663 Kaiserslautern, Germany}
\address[PCFT]{Peng Huanwu Center for Fundamental Theory, Hefei, Anhui 230026, China}
\address[ICTS]{Interdisciplinary Center for Theoretical Study, University of Science and Technology of China, Hefei, Anhui 230026, China}
\address[MPI]{Max-Planck-Institut f\"ur Physik,  Werner-Heisenberg-Institut, Boltzmannstraße 8, 85748 Garching, Germany}
\address[HU-IRIS]{Institut f\"ur Physik und IRIS Adlershof, Humboldt–Universität zu Berlin, 10099 Berlin, Germany}
\address[HU]{ Institut f\"ur Physik, Humboldt–Universität zu Berlin, Newtonstraße 15, 12489 Berlin, Germany}
\address[PKU]{Center for High Energy Physics, Peking University, Beijing 100871, People’s Republic of China}

\begin{abstract}
We introduce a new version v1.1 of {\NeatIBP}. In this new version, a {\Kira} interface is included. It allows the user to reduce the integration-by-parts (IBP) identity systems generated by {\NeatIBP} using {\Kira} in a highly automated way. This new version also implements the so-called \textit{spanning cuts} method. It helps to reduce the total computational complexity of IBP reduction for certain hard problems. Another important feature of this new version is an algorithm to simplify the solution module of the \textit{syzygy equations} hinted by the idea of \textit{maximal cuts}.
\end{abstract}


\end{frontmatter}

\noindent {\bf PROGRAM SUMMARY}\smallskip

\begin{small}
\noindent
{\em Manuscript Title:} {Performing integration-by-parts reductions using {\sc Kira}+{\sc NeatIBP} 1.1}\\
{\em Authors:} {Zihao Wu, Janko B{\"o}hm, Rourou Ma, Johann Usovitsch,  Yingxuan Xu, Yang
Zhang}\\
{\em Program Title:}   {NeatIBP 1.1}                                      \\
{\em CPC Library link to program files:} (to be added by Technical Editor) \\
{\em Developer's repository link:} \url{https://github.com/yzhphy/NeatIBP}\\
{\em Code Ocean capsule:} (to be added by Technical Editor)\\
{\em Licensing provisions(please choose one):} {GPLv3}  \\
{\em Programming language:}       {\sc Mathematica} 
         \\
{\em Computer(s) for which the program has been designed:} computers with multiple CPU cores\\
{\em Operating system(s) for which the program has been designed:} {Linux}   \\
{\em Supplementary material:} {none}                                \\
{\em Nature of problem:} Upgrading {\NeatIBP} \cite{1}, a package to decrease the difficulty of integration-by-parts reduction (IBP) of Feynman integrals via the syzygy methods.
\\
{\em Solution method:} We upgraded {\NeatIBP} to its new version to increase its capability and usability. We developed an interface to {\Kira} \cite{2,3}. With it, the user can reduce the linear system of IBP generated by {\NeatIBP} automatically through {\Kira}. We implemented the so-called \textit{spanning cuts} method in the new version. We also developed the syzygy simplification algorithm using the idea of \textit{maximal cut}. The latter two features increased the capability of {\NeatIBP} for solving harder Feynman integral families. Several other usability upgrades are also included in the new version. 
   \\

\end{small}

\section{Introduction}
\label{sec:intro}
In recent research in high energy physics, multi-loop Feynman integral computation becomes increasingly important. It is crucial for the evaluation of higher-order corrections of scattering amplitudes, which are critical for evaluating observables in high energy experiments. Nowadays, there is a growing demand for multi-loop Feynman integral computation, both for experimental and theoretical reasons. On the other hand, the computation of Feynman integrals is still challenging.

Currently, the Feynman integral computation forms a sophisticated work flow. We first generate the integrals from the Feynman diagrams \cite{Nogueira:1991ex,Vermaseren:2000nd, Kuipers:2012rf, Mertig:1990an, Shtabovenko:2016sxi, Shtabovenko:2020gxv, Shtabovenko:2023idz}. With the integrals, we use the tensor reduction methods \cite{Ossola:2006us,Ossola:2007ax,Passarino:1978jh} to reduce them to scalar integrals. After this, we apply integration-by-parts (IBP) reduction \cite{Tkachov:1981wb,Chetyrkin:1981qh} to reduce the scalar integrals to master integrals. Finally, we express the master integrals as analytic functions or numerical values \cite{Henn:2013pwa,Henn:2014qga,Liu:2017jxz,Liu:2018dmc,Guan:2019bcx,Zhang:2018mlo,Wang:2019mnn,Beneke:1997zp,Smirnov:2012gma,Lee:2017yex,Lee:2015eva,Lee:2009dh,Tarasov:1996br,Lee:2017ftw,peskin1995introduction,Lee:2013hzt,Lee:2013mka,Lee:2014tja,Baikov:1996cd,Baikov:1996rk,Baikov:2005nv,Smirnov:1999gc,Tausk:1999vh,Smirnov:2021rhf,Smirnov:2015mct,Smirnov:2013eza,Smirnov:2009pb,Smirnov:2008py,Smirnov:2023yhb,Borowka:2017idc,Borowka:2015mxa,Jahn:2018zsh}.

In the above steps, the integration-by-parts (IBP) reduction \cite{Tkachov:1981wb,Chetyrkin:1981qh} is one of the critical and challenging steps. By inducing linear relations from integration-by-parts formulae, the integrals relevant for a given scattering process, with a large amount, can be reduced as linear combinations of a smaller number of linearly independent integrals, called \textit{master integrals}. There are multiple programs for this purpose on the market \cite{Anastasiou:2004vj,Peraro:2019svx,Peraro:2016wsq,Peraro:2019okx,Smirnov:2008iw,Smirnov:2013dia,Smirnov:2014hma,Smirnov:2019qkx,Maierhofer:2017gsa,Maierhofer:2018gpa,Maierhofer:2019goc,Klappert:2020nbg,Usovitsch:2022wvr,Lange:2024mmz,Klappert:2019emp,Klappert:2020aqs,Guan:2024byi,Guan:2019bcx,Liu:2021wks,Magerya:2022hvj,Lee:2013mka,Studerus:2009ye,vonManteuffel:2012np,Boehm:2020ijp,Bendle:2021ueg,Heller:2021qkz}. However, the computational cost for IBP reduction has been a limitation for a long time. The traditional method of IBP reduction, the Laporta algorithm \cite{Laporta:2000dsw}, performs the reduction by solving a large system of linear equations through Gaussian elimination. Consequently, the efficiency of this method is limited by the number of linear equations in the system, which is usually a huge amount for frontier problems. Thus, the IBP reduction step is usually a bottleneck for these problems.

Recently, the development of \textit{syzygy} methods \cite{Gluza:2010ws,Chen:2015lyz,Larsen:2015ped,Zhang:2016kfo,Schabinger:2011dz,Bohm:2017qme,Bosma:2018mtf,Boehm:2020zig,Wu:2023upw} provides a new approach to provide a solution to this problem\footnote{There is a multitude of further IBP reduction methods, see \cite{Mastrolia:2018uzb,Frellesvig:2019uqt,Frellesvig:2019kgj,Frellesvig:2020qot,Smirnov:2006wh,Smirnov:2006vt,Smirnov:2006tz,Smirnov:2005ky,Bitoun:2017nre,Lee:2014tja,Baikov:1996iu,Lee:2008tj,Grozin:2011mt,Smirnov_2006,Smirnov:2012gma,vonManteuffel:2014ixa,Peraro:2016wsq,Klappert:2019emp,Klappert:2020aqs,Peraro:2019svx,Kosower:2018obg,Feng:2024qsa}.}. These methods introduce constraints on the propagator power indices, which leads to polynomial equations called \textit{syzygy equations}. Solving these equations effectively eliminates a large number of redundant integrals with propagator indices increased in the symbolic IBP identities, thus replacing the Gaussian elimination of these redundant integrals. Consequently, the size of the linear system of specific IBP identities needed in a reduction is much smaller compared to that from the traditional Laporta algorithm. This leads to a reduction with much higher efficiency. 

{\NeatIBP} \cite{Wu:2023upw} is a {\Mathematica} package which implements the syzygy methods. Specifically, it uses syzygy equations and module intersections in the Baikov representation \cite{Baikov:1996cd,Baikov:1996rk,Baikov:2005nv} to generate a small-sized IBP system for the user-input target integrals. Several examples in \cite{Wu:2023upw} show that the number of equations in the linear system is smaller by several degrees of magnitude compared to Laporta's algorithm. Thus, it is very helpful to solve the computational bottleneck in frontier problems. Until now, {\sc NeatIBP} has demonstrated its advantage by supporting a number of calculations for phenomenological or theoretical studies \cite{Feng:2024heh,Brancaccio:2024map,Abreu:2024yit,Badger:2024fgb, Badger:2024sqv, Liu:2024ont, Badger:2024awe, Badger:2024gjs}.

However, along with the advantages, there are also multiple areas that can be further improved in {\NeatIBP}. One limitation in earlier versions was that the package was only an IBP-system generator. It did not solve the system, as often required by the user. After running {\NeatIBP}, the user had to solve the system separately, either manually or with the help of relevant software, including {\sc Blade} \cite{Guan:2024byi,Guan:2019bcx,Liu:2021wks}, {\sc FiniteFlow} \cite{Peraro:2016wsq,Peraro:2019okx}, {\sc FireFly} \cite{Klappert:2019emp,Klappert:2020aqs}, {\sc Kira} \cite{Maierhofer:2017gsa,Maierhofer:2018gpa,Maierhofer:2019goc,Klappert:2020nbg,Usovitsch:2022wvr,Lange:2024mmz}, {\sc RaTracer} \cite{Magerya:2022hvj}, etc.

Consequently, to make {\NeatIBP} more convenient for users, it is highly beneficial to have an automatic IBP system solver with it. One natural idea is to develop an interface with the software mentioned above, to let them solve the system generated by {\NeatIBP} automatically. {\sc Kira} \cite{Maierhofer:2017gsa,Maierhofer:2018gpa,Maierhofer:2019goc,Klappert:2020nbg,Usovitsch:2022wvr,Lange:2024mmz}, as we mentioned before, is a widely used software for IBP reduction. And, it is very well suited for interfacing with {\sc NeatIBP}. Originally developed to implement the Laporta algorithm for reducing scalar multi-loop integrals, {\sc Kira} utilizes Feynman integral computations, employing modular arithmetic, direct Gaussian elimination techniques and finite field reconstruction methods. A major enhancement introduced in version 1.2 is the \textit{user-defined system} module, which allows users to input user-provided linear systems for reduction. This feature is very suitable to interface with {\sc NeatIBP}. By combining the IBP generation capabilities of {\sc NeatIBP} with the powerful equation-solving features of {\sc Kira}, researchers can tackle highly complex Feynman integral problems with improved efficiency and reduced computational cost. And, an interface will make this process highly automated.

In addition, several algorithmic improvements are of great advantage to increase the capabilities of {\NeatIBP}. One of the important ideas here is to generate IBP identities on \textit{generalized unitarity cuts} \spcref. The computation is much more effective on the cuts. Using this idea, the so-called \textit{spanning cuts} method performs IBP reduction on a set of cuts and merges the reduction results on the cuts. This method often decreases the computation cost and makes several previously intractable problems manageable. Another practical usage of the idea of cuts is to reduce redundancy in IBP syzygy vectors, which correspond to the generators of the module intersection, using the idea of \textit{maximal cuts}. These vectors generate IBP identities that contain only sub-sector integrals. Thus, they are removable in the \textit{tail mask} strategy. This simplification is beneficial for the efficiency of some problems.

In this paper, we introduce the new version v1.1 of {\sc NeatIBP}. It includes several significant updates and new features. One of the key enhancements is the integration of an interface with the software {\Kira} introduced above. Through this interface, {\sc NeatIBP} can automatically generate IBP systems, which are then directly solved by {\sc Kira}, streamlining the workflow.

Additionally, version 1.1 introduces the automated implementation of the \textit{spanning cuts} method. With this feature, IBP systems can be generated on different cut conditions, and the reduction results are automatically solved by {\sc Kira} and subsequently merged by {\sc NeatIBP}. The new features also include the mechanism to simplify the syzygy vectors through the \textit{maximal cut} idea.

This paper is organized as follows. In Section \ref{sec:review}, we give some reviews of the theory and some descriptions of the notation. In Section \ref{sec: features}, we describe the new features in version 1.1, as well as the corresponding algorithms. Section \ref{sec:manual} is the manual on how to use the new features. In Section \ref{sec: examples}, we give some examples. In Section \ref{sec: summary}, we summarize this paper.

We remark that due to application requirements in universities in China, and Wu's significant contribution to this work, the author name ordering of this paper is chosen in the stated way.

\section{Review of methods and notations}\label{sec:review}
\subsection{IBP reduction from Baikov representation and the syzygy equations}
In Baikov representation, the Feynman integrals are expressed as
\begin{equation}
    I_{\alpha_1,\cdots,\alpha_n}=C\int_\Omega \dif z_1\cdots \dif z_n\frac{P^\gamma}{{z_1}^{\alpha_1}\cdots {z_n}^{\alpha_n}},
\end{equation}
where $C$ and $\gamma$ are constants, and $\Omega$ is the integration region with a boundary where $P=0$. 

The IBP identities of Baikov representations can be generated by 
\begin{equation}
    0=C\int_\Omega \dif z_1\cdots \dif z_n\sum_{i=1}^n\frac{\partial}{\partial z_i}\frac{a_i(z)P^\gamma}{{z_1}^{\alpha_1}\cdots {z_n}^{\alpha_n}},
\end{equation}
If the polynomials $a_i$ are chosen such that the following equations are satisfied,
\begin{equation}\label{eq:syzygy equations}
\begin{cases}
    
    \sum_{i=1}^n \Big(a_i(z)\frac{\partial P}{\partial z_i}\Big)+ b(z) P =0, \\
    a_i(z)-b_i(z) z_i=0, \quad \text{for } i\in \{j|\alpha_j>0\},
    \end{cases}
\end{equation}
we get IBP identities free of dimension shift and increasing of denominator indices, as follows
\begin{equation}\label{eq:Baikov IBP}
    0=C\int\dif z_1 \cdots \dif z_n \Big(\sum_{i=1}^n\big( \frac{\partial a_i}{\partial z_i} -\alpha_i \frac{ a_i}{ z_i}\big)-\gamma b\Big) P^\gamma \frac{1}{z_1^{\alpha_1} \cdots z_n^{\alpha_n}}.
\end{equation}
The set $S:=\{j|\alpha_j>0\}$ labels a \textit{sector} that the corresponding integrals belong to. 

In the language of algebraic geometry, equations \eqref{eq:syzygy equations} are a set of \textit{syzygy equations}. Their solutions form a \textit{module}. Solving the equations is to find a set of \textit{generators} of the solution module. Then, each generator $(a_i,b)$ corresponds to an IBP identity through \eqref{eq:Baikov IBP}. Such an identity is considered as a \textit{formal IBP relation} as long as the $\alpha_i$ are interpreted as symbolic indices. Taking them as specific numbers, we get specific IBP relations. This process is called \textit{seeding}, and the tuple of the set of numbers is called a \textit{seed}. 

\subsection{Generalized unitarity cut and spanning cuts}\label{sec: cut}
The \textit{(generalized unitarity) cut} \spcref is an important concept for Feynman integral computation, and it is extensively adopted in {\NeatIBP}. In Baikov representation, cutting an integral is to replace the integration over certain $z_i$'s by taking their residues, i.e.
\begin{equation}\label{eq: C-cut}
    I_{\alpha_1,\cdots,\alpha_n}|_{\C\text{-cut}}:=C\int_{\Omega_\C}\prod_{i\notin\C} \dif z_i\prod_{i\in\C}\oint_0 \dif z_i\frac{P^\gamma}{{z_1}^{\alpha_1}\cdots {z_n}^{\alpha_n}},
\end{equation}
where $\C$ is a subset of indices.

The idea of cut is useful to simplify IBP reduction, because after cut many integrals vanish\footnote{More specifically, if an integral $I$ is in a sector labeled by $S$, then we have $I|_{\C\cut}=0$ if $\C\not\subseteq S$.}, and it is observed that this operation preserves IBP relations in many cases.

This concept allows us to use the so-called \textit{spanning cuts}\footnote{See \cite{Bohm:2018bdy,Bendle:2019csk,Boehm:2020zig,Bern:2024adl} for several works that apply this idea.} method to simplify the IBP reduction. The idea is to find a set of spanning cuts $\{\C_i\}$ such that, for any non-zero sector $S$ in a family, there exists at least one $\C\in\{\C_i\}$, such that $\C\subseteq S$. A usual choice of spanning cuts is the set of all bottom sectors. Then, we perform IBP reduction on each cut in $\{\C_i\}$. Due to the definition of the spanning cuts, the reduction coefficients in front of a given master integral shall be detected at least once on a certain cut. Thus, we can finally collect the coefficients on all of the cuts and merge them to a complete reduction result. The spanning cuts method converts a single, large-scaled reduction problem to several smaller problems. This, in many cases, makes the computation easier and allows for parallelization.

\section{Main new features }\label{sec: features}

\subsection{The {\sc NeatIBP+Kira} interface}\label{sec: feature kira interface}
In its earlier versions, {\NeatIBP} serves as an IBP identity generator without reducing module. Thus, of course, there is demand for an automated reducer for better use of the package. In {\NeatIBP} v1.1, we include an interface with {\Kira}. It converts {\NeatIBP}'s output into a format compatible with \Kira, and allows users to leverage {\Kira} for the reduction of linear systems. Since the release of {\Kira} 1.2 \cite{Maierhofer:2018gpa}, {\Kira} has offered a robust option to reduce user-defined systems.
In this module, the user-defined system of equations to be reduced can be imported directly without requiring the {\Kira} config files \texttt{integralfamilies.yaml} and \texttt{kinematics.yaml}. 
For detailed usage and settings for reducing user-defined systems, please refer to Refs.~\cite{Maierhofer:2018gpa, Klappert:2020nbg}, as well as the examples provided in the {\Kira} package as
\begin{lstlisting}
kira/examples/userDefinedSystem1
\end{lstlisting}
and
\begin{lstlisting}
kira/examples/userDefinedSystem2
\end{lstlisting}

The interface to {\Kira}'s user-defined system in {\NeatIBP} can be activated by a few simple settings in the {\NeatIBP} input file \texttt{config.txt}\footnote{See Section \ref{sec: kira interface manual} for the corresponding manual.}. It performs the following procedures automatically. Firstly, it converts the IBP system generated by {\NeatIBP} to {\Kira} readable format. During conversion, the interface sorts the IBP identities according to the criteria required by {\Kira}. The criteria are:
\begin{enumerate}
    \item Compare the most complicated\footnote{Being complicated here means to be less preferred by the integral ordering.} integrals appearing in the IBPs. The more complicated the integral is, the less preferred the corresponding IBP is. 
    \item If the above criterion does not distinguish some of the IBPs, compare the number of integrals appearing in the IBPs. The more integrals appear, the less preference the corresponding IBPs will have.
    \item If the above criteria do not distinguish some of the IBPs, compare the second most complicated integrals appearing in the IBPs. The more complicated the integral is, the less preferred the corresponding IBP is.
    \item If the above criteria do not distinguish some of the IBPs, compare the third complicated integral appearing in the IBPs.
    \item And so on...
    
\end{enumerate}
After the conversion, the interface reduces the system using a \textit{user-defined system} in {\Kira}. After {\Kira} finishes, the interface collects and converts the result to a {\sc Mathematica} readable file.

\subsection{The implementation of the spanning cuts method}\label{sec:spc}
In {\NeatIBP} v1.1, the spanning cuts method introduced in Section \ref{sec: cut} is implemented. When activating the spanning cuts mode in the configuration file of {\NeatIBP}, {\NeatIBP} first finds a set of spanning cuts. Usually, the choice is the bottom sectors. {\NeatIBP} does so when the target integrals are without multiple propagators. If there are some target integrals with multiple propagators, because {\NeatIBP} does not support cutting multiple propagators, yet, it will find a different choice of spanning cuts that avoids cutting the multiple propagators. The strategy is as follows.
\begin{enumerate}
    \item Determine all bottom sectors, set spanning cuts $\{\C\}$ as the set of bottom sectors.
    \item Delete the index where multiple propagator(s) appear in each cut in $\{\C\}$.
    \item Delete duplicated members in $\{\C\}$.
    \item Delete the cut in $\{\C\}$ such that there exists another cut in $\{\C\}$, which is its subset.
\end{enumerate}
For example, if the set of bottom sectors are $\{\{1,4,6\},\{1,4,7\},\{2,5,7\}\}$, and there appears a multiple propagator on index $7$, the spanning cuts will be $\{\{1,4\},\{2,5\}\}$\footnote{This notation means that the spanning cuts consist of two different choices of cut. They are $\{1,4\}\cut$ and $\{2,5\}\cut$.}. 

One extreme example is that there is no index which is free of multiple propagators. In this case, the spanning cuts will be $\{\{\}\}$\footnote{This notation means that the spanning cuts consist of one choice of cut. It is $\{\}\cut$, which means not cutting any propagators at all.}. This is equivalent to not cutting at all. To avoid this, the user is recommended to separate the target integrals categorized by the position of multiple propagators, and to run spanning cuts individually on each category. Then, in each category, the number of indices that cannot be cut will be much smaller.



\begin{figure}[htb]
\centering
\begin{tikzpicture}[node distance=3cm,scale=0.5]
\node (0) [process] {Determine spanning cuts};
\node (c1) [process,yshift=-2cm,xshift=-4cm ] {Generate IBPs on $\C_1\cut$};
\node (c2) [process,yshift=-2cm,xshift=0cm ] {Generate IBPs on $\C_2\cut$};
\node (cn) [plaintext,yshift=-2cm,xshift=4cm ] {...};
\node (ccheck) [process,yshift=-4cm,xshift=0cm ] {Consistency Check};
\node (s) [process,yshift=-6cm,xshift=0cm ] {Shorten IBPs (optional)};
\node (k1) [process,yshift=-8cm,xshift=-4cm ] {Reduce IBPs on $\C_1\cut$};
\node (k2) [process,yshift=-8cm,xshift=0cm ] {Reduce IBPs on $\C_2\cut$};
\node (kn) [plaintext,yshift=-8cm,xshift=4cm ] {...};
\node (merge) [process,yshift=-10cm,xshift=0cm ] {Merge reduced IBPs on all cuts};
\draw [arrow] (0) -- (c1);
\draw [arrow] (0) -- (c2);
\draw [arrow] (0) -- (cn);
\draw [arrow] (c1) -- (ccheck);
\draw [arrow] (c2) -- (ccheck);
\draw [arrow] (cn) -- (ccheck);
\draw [arrow] (ccheck) -- (s);
\draw [arrow] (s) -- (k1);
\draw [arrow] (s) -- (k2);
\draw [arrow] (s) -- (kn);
\draw [arrow] (k1) -- (merge);
\draw [arrow] (k2) -- (merge);
\draw [arrow] (kn) -- (merge);
\end{tikzpicture}
\caption{The work flow of spanning cuts method using {\NeatIBP}+{\Kira}.}
\label{fig: work flow spc}
\end{figure}
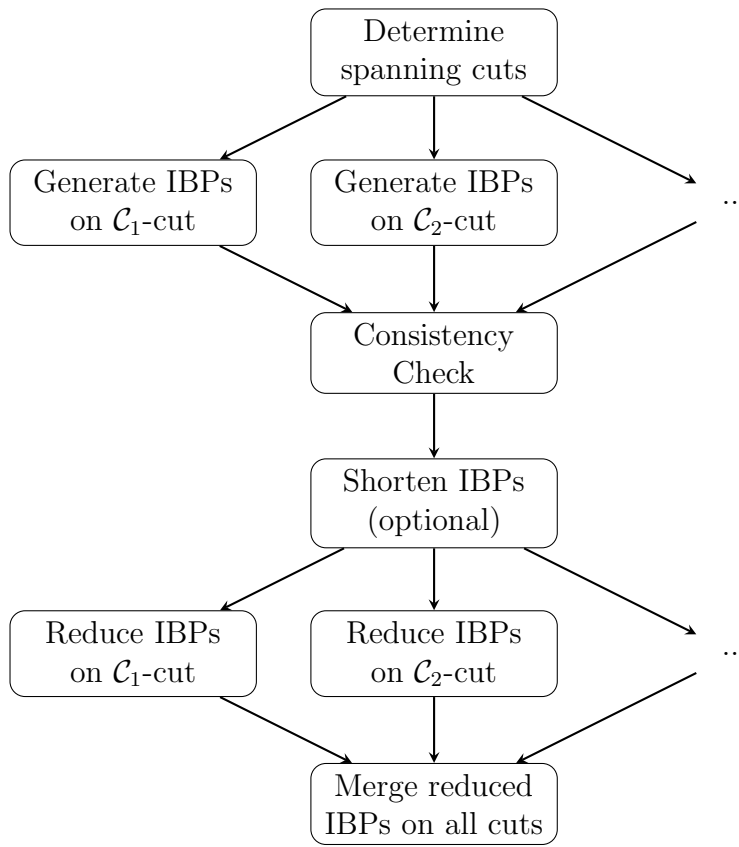

The spanning cut mode, running together with the {\Kira} interface, forms a complete work flow of the spanning cuts method. It is shown in Fig. \ref{fig: work flow spc}.

In Fig. \ref{fig: work flow spc}, the step consistency check is checking the \textit{consistency condition} for the spanning cuts IBPs. This condition says that, if an IBP reduction coefficient is detected in more than one cut, the detected results are expected to be equal. If the consistency check fails, {\NeatIBP} will not continue the remaining steps. This condition is checked in {\NeatIBP} using a numerical reduction on finite fields. The numerical point is chosen as the \texttt{GenericPoint}, \texttt{GenericD} in the {\NeatIBP} input files, and the finite field modulus is chosen as \texttt{FiniteFieldModulus} in the input files.

The consistency check is essential because this condition, as we observed, is not always guaranteed. Our observations indicate that diagrams with massive propagators are more prone to inconsistencies. Although the exact cause is not yet clear, it is more likely related to theoretical or algorithmic factors, rather than any bugs in the codes. To make this claim more convincing, we state that such inconsistencies can also be produced in {\Kira}, individually without {\NeatIBP}. In {\Kira}, when an IBP system is generated by setting integrals belonging to cut sectors to zero before any reduction step, new relations arise from truncated equations. These truncated equations can sometimes relate integrals inconsistently. Moreover, we have checked the IBP identities generated by {\NeatIBP} in some of the inconsistent cases, aligned with those produced by {\Kira} when used independently, without {\NeatIBP}. It turns out that the IBP identities generated by {\NeatIBP} indeed vanishes using pure {\Kira} reduction.

However, we claim that, the inconsistency issue does not invalidate the implementation of the spanning cuts method in the current version of {\NeatIBP}. Firstly, as stated above, in many cases, where the diagrams are without massive propagators, the consistency condition is satisfied. For these diagrams, the spanning cuts method appears to be very powerful and useful. Secondly, the implementation of the spanning cuts method helps us to study the reason of the inconsistency problem more efficiently, which is our ongoing work. Thirdly, once the root cause of the inconsistency is identified and the corresponding algorithm adjustments are made in {\NeatIBP}, the idea of spanning cuts will hopefully become adoptable to diagrams with massive propagators as well.

After the consistency check, if it is passed, there is an optional IBP shortening step. In this step, if a master integral is shared by two or more cuts, we can pick up one simplest cut, and set the corresponding master integral to zero in the IBP identities generated on all the other cuts. Since the consistency condition holds, it is safe to do so. This makes the system smaller and makes the reduction easier.

The next steps are reducing the IBP systems on each cut. Currently, these reductions are performed by {\Kira} through the interface introduced in Section \ref{sec: feature kira interface}. After getting the reduced results, {\NeatIBP} uses a direct strategy to merge them. Let $I$ to be the target integral and $I_i$ to be the master integrals, and on $\C_i\cut$ the reduced result is
\begin{equation}
    I=\sum_j c^{\C_i\cut}_j I_j,
\end{equation}
where $c^{\C_i\cut}_j$ are the IBP reduction coefficients on different cuts. We label the merged IBP reduction result as
\begin{equation}
    I=\sum_j c_j I_j.
\end{equation}
For a given $j$, the algorithm of merging strategy is described as follows:
\begin{enumerate}
    \item For all $i$ such that $I_j$ should vanish on $\C_i\cut$, check if we have $c^{\C_i\cut}_j=0$. If not, return \texttt{\$Failed}.
    \item If the reduction results are from the shortened systems, pick up all $c^{\C_i\cut}_j$ \textbf{that do not vanish}, and check if they equal to each other. If so, select one of them and let $c_j$ to be it. If not, return \texttt{\$Failed}. 
    \item If the reduction results are \textbf{not} from the shortened systems, pick up all $c^{\C_i\cut}_j$ \textbf{such that $I_j$ shall not vanish on $\C_i\cut$}, and check if they equal to each other. If so, select one of them and let $c_j$ to be it. If not, return \texttt{\$Failed}. 
\end{enumerate}

We have two comments on the above algorithm. Firstly, in the above algorithm, we need to check whether a coefficient vanishes or whether two coefficients are equal. These checks are performed numerically on rational numbers. Secondly, the appearance of \texttt{\$Failed} in the above algorithm means that the results of IBP reduction on the cuts are not consistent. This will not possibly happen if the consistency check step is passed. Even though, we remind that the consistency check step is a numerical check, there is indeed a very small chance that it makes mistakes because of the badly chosen numeric point. Thus, the random numerical checks in the above algorithms also serve as double checks.

\subsection{Simplifying the syzygy vectors using the idea of maximal cut}\label{sec: syzygy simplification algorithm}

In this sub section, we discuss the new simplification algorithm of the syzygy generators implemented in the new version of {\NeatIBP}. The main concern is the number of generators in the solution module of the syzygy equations \eqref{eq:syzygy equations}. For some examples, the number of generators can easily be up to hundreds or even thousands. This is a heavy burden for the subsequent steps, including the generation of formal IBP relations and the seeding. For the new version of {\NeatIBP}, we developed an algorithm to decrease the number of generators by deleting the redundant ones. We discuss this algorithm in two cases in the following two sub-subsections.

\subsubsection{For cases without multiple propagators}\label{sec: mc simp no mp}
Notice that, since {\NeatIBP} employs the so-called \textit{tail mask} strategy, the IBP relations that only contain integrals from the sub-sectors (called \textit{ tail integrals}) will be discarded. This allows us to delete a large number of the generators, which corresponds to such sub-sector IBP relations. In practice, consider the formal IBP identity \eqref{eq:Baikov IBP}. It can be rewritten as
\begin{equation}
    0=C\int\dif z_1 \cdots \dif z_n  \xi(a_i,b,\{\alpha_i\}) \frac{P^\gamma}{z_1^{\alpha_1} \cdots z_n^{\alpha_n}},
\end{equation}
where
\begin{equation}
    \xi(a_i,b,\{\alpha_i\})=\sum_{i\in S}\big( z_i\frac{\partial b_i}{\partial z_i} -(\alpha_i-1) b_i\big)+\sum_{i\notin S}\big( \frac{\partial a_i}{\partial z_i} -\alpha_i \frac{ a_i}{ z_i}\big)-\gamma b,
\end{equation}
and $S$ is the sector. 

For cases where the integrals to be reduced are without multiple propagators, we usually (though there could be counterexamples) do not need seeds with indices larger than 2 in the seeding step. That is, for $i\in S$, we have $\alpha_i=1$. In these cases, that an IBP relation contains only tail integrals is equivalent to
\begin{equation}
    0=\xi(a_i,b,\{\alpha_i\})|_{\mc},
\end{equation}
where ``mc'' stands for \textit{maximal cut}, meaning to set $z_i=0$ for all $i \in S$, which is the way we apply an $S$-cut for integrals without multiple denominators in sector $S$. For cases that $\alpha_i=1$ for $i \in S$, we have
\begin{equation}
\begin{aligned}
    &\xi(a_i,b,\{\alpha_i\})|_{\mc}\\
    &=\Big(\sum_{i\notin S}\big( \frac{\partial a_i}{\partial z_i} -\alpha_i \frac{ a_i}{ z_i}\big)-\gamma b\Big)\Big|_\mc\\
   &=\sum_{i\notin S}\big( \frac{\partial a_i|_\mc}{\partial z_i} -\alpha_i \frac{ a_i|_\mc}{ z_i}\big)-\gamma b|_\mc
\end{aligned}
\end{equation}
Thus, a generator $(a_i,b)$ corresponds to IBP identities with only tail integrals if it satisfies $(a_i,b)|_\mc=0$\footnote{Notice that $a_i|_\mc=0$ always holds for $i\in S$, because we have $a_i=z_i b_i$ for $i\in S$.}. This allows us to simplify the module $M$ generated by these generators, by finding the subset of the generators such that deleting these generators does not change the maximal-cut module. Specifically, let
\begin{equation}
    M=\langle f_1,\cdots,f_m\rangle
\end{equation}
with $f_j$'s the generator vectors. We define the maximal-cut module as 
\begin{equation}
    M|_\mc=\langle f_1|_\mc,\cdots,f_m|_\mc\rangle.
\end{equation}
We delete $f_j$ in the generating set if this does not change $M|_\mc$, i.e.
\begin{equation}
    M|_\mc=\langle f_1|_\mc,\cdots,f_{j-1}|_\mc,f_{j+1}|_\mc\cdots,f_m|_\mc\rangle.
\end{equation}
The algorithm is as follows. Firstly, we apply the algorithm ``lift selection'', shown in Algorithm \ref{algorithm: lift selection}.
\begin{algorithm}[htpb]
  \SetAlgoLined
  \KwIn{generating set $F=\{f_i\}$}
  Sort $F$ by preference, to make the preferred generators are more likely to survive in the following selection.\\
  {Compute the Gr{\" o}bner basis $G|_\mc$ of $\{f_i|_\mc\}$, with generators }$\{g_j\}$.\\
  Lift the generators: $g_j=\sum_kc_{jk}f_k|_\mc$.  \\
  \KwOut{The selected generators $\{f_k|\exists j, c_{jk}\neq0\}$}
  \caption{lift selection}
  \label{algorithm: lift selection}
\end{algorithm}
Using the lift selection algorithm, we select the generators that contribute to the Gr{\" o}bner basis of maximal-cut module, and delete those that do not. After this step, we can apply a ``further selection'' shown in Algorithm \ref{algorithm: further selection}.
\begin{algorithm}[htpb]
  \SetAlgoLined
  \KwIn{generating set $F=\{f_i\}$}
  Sort $F$ by preference. The preferred generators are put in the front.\\
  $G|_\mc=$Gr{\" o}bnerBasis$(F|_\mc)$\\
  \For{$i$ \tn{from} $m$ \tn{to} $1$}{
    $F^\prime=$ delete $f_i$ from $F$\\
    $G^\prime|_\mc=$Gr{\" o}bnerBasis$(F^\prime|_\mc)$\\
    \If{$G^\prime|_\mc$ == $G|_\mc$}{
        $F=F^\prime$
    }
  }
  \KwOut{The selected generators $F$}
  \caption{further selection}
  \label{algorithm: further selection}
\end{algorithm}

The Gr{\" o}bner basis and lift computations in Algorithm \ref{algorithm: lift selection} and \ref{algorithm: further selection} are performed numerically over a finite field. The numerical point is \texttt{GenericPoint} and \texttt{GenericD}, and the finite field modulus is \texttt{FiniteFieldModulus2}, which can be defined in the user's input file(s). Note that \texttt{FiniteFieldModulus2} is specifically for use in these algorithms.

\subsubsection{For cases with multiple propagators}\label{sec: mc simp with mp}
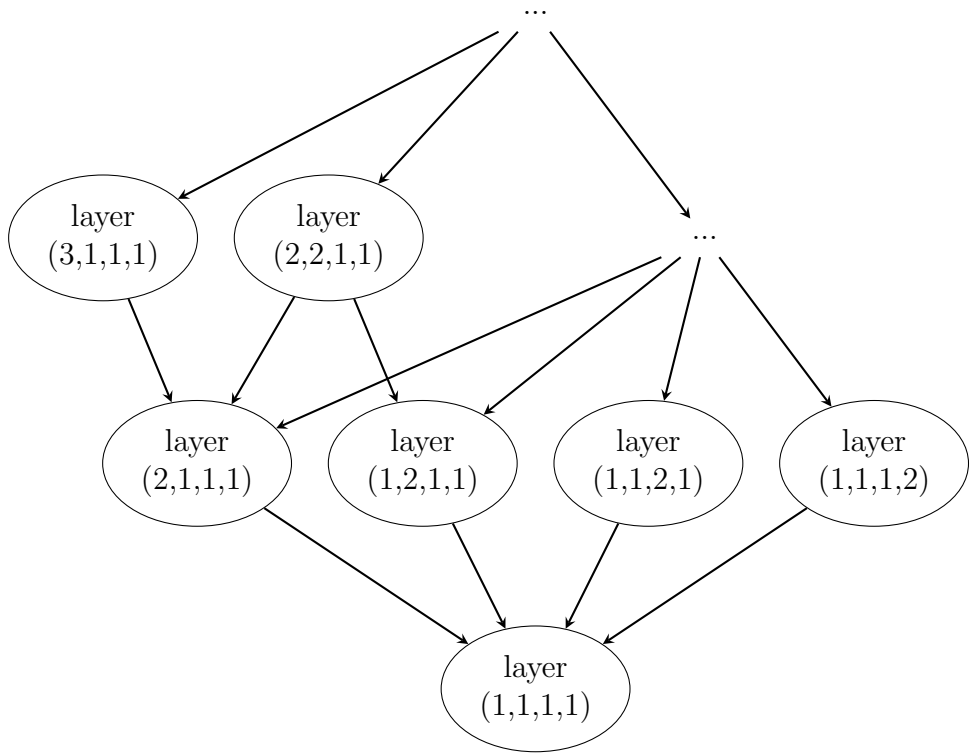
\begin{figure}[htb]
\centering
\begin{tikzpicture}[node distance=3cm,scale=0.5]
\node (1111) [item] {layer (1,1,1,1)};
\node (2111) [item,yshift=3cm,xshift=-4.5cm ] {layer (2,1,1,1)};
\node (1211) [item,yshift=3cm,xshift=-1.5cm ] {layer (1,2,1,1)};
\node (1121) [item,yshift=3cm,xshift=1.5cm ] {layer (1,1,2,1)};
\node (1112) [item,yshift=3cm,xshift=4.5cm ] {layer (1,1,1,2)};
\node (3111) [item,yshift=6cm,xshift=-5.75cm ] {layer (3,1,1,1)};
\node (2211) [item,yshift=6cm,xshift=-2.75cm ] {layer (2,2,1,1)};
\node (others) [plaintext,yshift=6cm,xshift=2.25cm ] {...};
\node (others2) [plaintext,yshift=9cm,xshift=0cm ] {...};
\draw [arrow] (others2) -- (2211);
\draw [arrow] (others2) -- (3111);
\draw [arrow] (others2) -- (others);
\draw [arrow] (others) -- (2111);
\draw [arrow] (others) -- (1211);
\draw [arrow] (others) -- (1121);
\draw [arrow] (others) -- (1112);
\draw [arrow] (3111) -- (2111);
\draw [arrow] (2211) -- (2111);
\draw [arrow] (2211) -- (1211);
\draw [arrow] (1112) -- (1111);
\draw [arrow] (1121) -- (1111);
\draw [arrow] (1211) -- (1111);
\draw [arrow] (2111) -- (1111);

\end{tikzpicture}
\caption{An example of a web structure of the layers in a sector with 4 denominators. The integrals with no multiple propagators are in layer (1,1,1,1). The integrals like $I_{2,1,1,1,-a,-b,\cdots}$ are in layer (2,1,1,1).}
\label{fig: layers}
\end{figure}

For cases with multiple propagators, we do not use the tail mask idea to design the algorithms. We use a generalization of it. Similar to that a Feynman integral family can be divided into sectors categorized by their denominator members, we can further divide a sector into different \textit{layers} by their denominator indices. For example, for a sector with 4 denominators, the layers are structured as shown in Fig. \ref{fig: layers}

Like the tail mask strategy, we adopt a similar ``layer mask'' strategy, considering that an integral in a certain layer can be reduced to simpler integrals. These integrals, if we forget about the sub-sector integrals, are distributed in the target integral's current layer and its sub-layers. We assume that, firstly, this reduction does not need seeds with denominator indices higher than the layer indices, and secondly, the sub-layer integrals can also be reduced to master integrals in the same way. With these assumptions, we can say that \textit{an IBP relation contains only sub-layer integrals can be deleted}.

Compared to cases without multiple propagators, $\alpha_i=1$ for $i\in S$ is not true for non-bottom layers. For these layers, we have 
\begin{equation}
\begin{aligned}
    &\xi(a_i,b,\{\alpha_i\})|_{\mc}\\
    &=\Big(\sum_{i\in S}(1-\alpha_i) b_i+\sum_{i\notin S}\big( \frac{\partial a_i}{\partial z_i} -\alpha_i \frac{ a_i}{ z_i}\big)-\gamma b\Big)\Big|_\mc\\
   &=\sum_{i\in S}(1-\alpha_i) b_i|_\mc+\sum_{i\notin S}\big( \frac{\partial a_i|_\mc}{\partial z_i} -\alpha_i \frac{ a_i|_\mc}{ z_i}\big)-\gamma b|_\mc
\end{aligned}
\end{equation}

Now the condition that makes $\xi(a_i,b,\{\alpha_i\})|_{\mc}=0$ becomes $(\tilde{a}_i,b)|_\mc=0$, where we define
\begin{equation}
    \tilde{a}_i=\begin{cases}
        b_i,\quad i\in S\\
        a_i,\quad i\notin S
    \end{cases}
\end{equation}

With the above, we consider the module $\title{M}$ generated by the generators $(\tilde{a}_i,b)|_\mc=0$. We delete those generators such that after the deletion (the Gr{\" o}bner basis of ) $\tilde{M}$ remains unchanged. This idea is almost the same as what we have in Section \ref{sec: mc simp no mp}. Thus, the algorithm we use is also the same as Algorithm \ref{algorithm: lift selection} and Algorithm \ref{algorithm: further selection}. The only difference is in the expressions of the generators.

We note that, in this case, the notation $|_\mc$ still refers to taking $\alpha_i=0$ for all $i\in S$. Thus, in such cases, this notation is no longer directly related to the concept \textit{generalized unitarity cut} (because we need to take derivative when computing residues for double poles). However, since we are already used to the phrase ``cut'' when describing the corresponding algorithms in the non-multiple-denominator cases, we continue to ``borrow'' this phrase in the generalized cases. We hope that this does not cause confusion for the users.

\subsubsection{Some comments}\label{sec:syzygy simplification comments}

Note that, although the simplification algorithms decrease the number of generators, thus lightens the burden of the formal IBP generation and the seeding, we do not claim that making use of the algorithms is always more efficient. The reason is that running the algorithms themselves takes time. In Section \ref{sec: syzygy simplification manual}, we introduce some instructions for introducing some time constraints in these algorithms.

Moreover, the simplification algorithms introduced above usually delete a large percentage of generators that are not needed. However, for some reason, the user might want to keep some of the unneeded generators. These unneeded generators sometimes serve as a ``catalyst'' that makes the seeding step even better performed. Thus, a ``strictness'' setting is included in the algorithms in {\NeatIBP}. By setting the strictness number to a value less than 1, {\NeatIBP} randomly\footnote{As a consequence, setting this value may make the outputs of the program to become different with same inputs.} keeps some of the generators that are supposed to be deleted. The probability of keeping a certain unneeded generator is $(1-\text{strictness})$. The way of setting the corresponding numbers of strictness will be introduced in Section \ref{sec: syzygy simplification manual}.

Notice that, the simplification in Section \ref{sec: mc simp no mp} is stronger than that of \ref{sec: mc simp with mp}. This means that the generators deleted by the latter one is a subset of the former. {\NeatIBP} determines which to use by the target integrals. If the target integrals are without multiple propagators, it chooses the former, otherwise the latter. There is a way to let {\NeatIBP} use the latter without regarding the target integrals. See Section \ref{sec: syzygy simplification manual} for corresponding settings.

We also note that, in {\NeatIBP}, there is a less sophisticated mechanism to remove sub-sector IBPs from its earlier versions. This method, named \texttt{NCornerKill} in the codes, looks into the formal (symbolic) IBPs\footnote{We remind the reader that each syzygy vector can be translated to a formal IBP.}, finds out symbolic integrals that appear in each formal IBP, and seeds them using \textit{numerator-corner}\footnote{This means there are no negative indices in the seeds.} seeds with the highest denominator indices appearing in the target integrals. If the seeded integrals are all with sub-sector integrals, the corresponding formal IBP will be deleted. With the new simplification algorithms introduced in the new versions of {\NeatIBP}, the older \texttt{NCornerKill} is no longer necessary. It is still on by default. But it can be turned off. We state the corresponding setting in Section \ref{sec: syzygy simplification manual}.

Another issue is that the methods introduced above to remove sub-sector IBPs are based on assumptions. As assumed, to generate a system sufficient to reduce given target integrals, we do not need seeds with denominator indices higher than the target integrals. Although in most cases this is true, there are some counterexamples. As stated in \cite{Wu:2023upw}, to deal with these counterexample cases, there is a module \texttt{AdditionalIBPs} in {\NeatIBP} to search for seeds with increased denominator indices. In this module, since the assumptions for these methods are no longer true, {\NeatIBP} uses the syzygy vectors (or say, formal IBPs) before applying the simplification algorithms.

\subsection{Other useful new features}\label{sec: other feature}

In this subsection, we outline some further useful new features. 

One of the features is the \textit{flexible degree bound} of the intersection computation in \Singular. For some hard problems, the {\Singular} computation of the module intersection could be a bottleneck that stalls the whole computation. Usually, this happens when user sets a relatively high degree bound in {\Singular}. In the new version of {\NeatIBP}, the user can introduce a constraint on the time used in the {\Singular} intersection computation. If the time exceeds this limit, {\Singular} terminates and runs again with the degree bound decreased by 1. If it still exceeds the time constraint, {\Singular} will rerun again with degree bound further decreased by 1. This loops until it finishes the computation in time, or the decreased degree bound reaches the user-appointed limit. In the latter case, {\NeatIBP} will report a failure for this sector. This mechanism is currently only available in the module intersection step for deriving formal IBPs. It is not included in the {\sc Azuritino} (see \cite{Wu:2023upw,Georgoudis:2016wff}) part for deriving master integrals. Because this part is usually not the bottleneck. 

We need to remind the reader that the parameters, including the time constraint and the degree-decreased limit, should be tuned to get a better performance. The instructions on this will be given in Section \ref{sec:singular related instructions}. Also, the trade-off using a lower degree bound for a faster {\Singular} computation is not always a good choice. If the degree bound in {\Singular} is too low, it is very likely that we have to frequently use denominator-increased seeds in the seeding steps, using the module \texttt{AdditionalIBPs} introduced in \cite{Wu:2023upw}. This will make the seeding step much more costly, both for time and memory.

In the new version, we also allow the user to specify the {\Singular} monomial ordering according to the usual notation of {\Singular}: 
\begin{center}
\url{https://www.singular.uni-kl.de/Manual/4-4/sing_985.htm}
\end{center}
The corresponding instructions are given in Section \ref{sec:singular related instructions}.

The new version also includes updates of the symmetry algorithms introduced in \cite{Wu:2024paw}. We will list some related settings in Section \ref{sec: symmetry manual}.

\section{Manual}\label{sec:manual}
This section is the manual for the new features introduced above. For users who are beginning to use {\NeatIBP}, they can download it using
\begin{lstlisting}[language=bash]
git clone https://github.com/yzhphy/NeatIBP.git
\end{lstlisting}
and follow the ``README'' or the manual in \cite{Wu:2023upw} for how to use the basic functions of this package. For people who already have this package, since the description in this paper is based on \version, if he or she has earlier versions, it is recommended to update it using
\begin{lstlisting}[language=bash]
git pull
\end{lstlisting}
See the ``README.md'' document in your local {\NeatIBP} files for your current version. For future versions, please refer to the ``README.md'' document 
at \url{https://github.com/yzhphy/NeatIBP} for possible updates of the manual, or refer to ``\texttt{developement\_documents/version_history.txt}'' for modification records of the historical versions.

\subsection{How to use the {\Kira} interface in {\NeatIBP}}\label{sec: kira interface manual}
To utilize the {\Kira} interface in {\NeatIBP}, it is necessary to have {\Kira} installed on your system. For detailed installation instructions, please refer to \cite{Maierhofer:2017gsa,Maierhofer:2018gpa,Maierhofer:2019goc,Klappert:2020nbg}. The installation files and guidelines are available at the following link:
\href{https://gitlab.com/kira-pyred/kira}{\texttt{https://gitlab.com/kira-pyred/kira}}.

Additionally, {\Kira} depends on the symbolic algebra program {\sc Fermat} \cite{fermat_software}. You can download Fermat and find further information about it at \url{http://home.bway.net/lewis/}.

To enable this interface, the user should include the following settings to his or her \texttt{config.txt} of the {\NeatIBP} input files:
\begin{lstlisting}[language=Mathematica]
PerformIBPReduction = True;
IBPReductionMethod = "Kira";
KiraCommand = "/some/path/kira";
FermatPath = "/some/path/fer64";
\end{lstlisting}

With these settings, when {\NeatIBP} finishes IBP generation, it converts the {\NeatIBP} outputs to {\Kira} readable input files, putting them into the {\Kira} reduction working folder as
\begin{lstlisting}
your_NeatIBP_outputPath/KiraIO/
\end{lstlisting}
Recall that during conversion, the interface sorts the IBP identities according to the criteria introduced in Section \ref{sec: feature kira interface}.

After the conversion, the interface calls {\Kira} to reduce the IBP system using the mechanism \textit{user-defined system} in {\Kira}. After the reduction, the interface reads the {\Kira} outputs and save the reduction results at 
\begin{lstlisting}
your_outputPath/results/Kira_reduction_results/reduced_IBP_Table.txt
\end{lstlisting}
It is a {\sc Mathematica} readable file as a list of replacement rules. The {\Kira} data base file \texttt{KiraIO/results/kira.db} in the {\Kira} working folder will be deleted after {\Kira} reduction finishes in order to save the disk space. If the user wants to keep it, set the following in \texttt{config.txt}
\begin{lstlisting}[language=Mathematica]
DeleteKiraDB = False;
\end{lstlisting}

The IBP solver for {\NeatIBP}, currently referring to {\Kira}, reduces the system analytically. If the user wants to do this numerically or partial numerically, set the following \Mathematica replacement rules in \texttt{config.txt}
\begin{lstlisting}[language=Mathematica]
NumericsForIBPReduction = {var1->xxx, var2->xxx, ... };
\end{lstlisting}

In addition, the \texttt{KiraCommand} setting not only specifies the location of {\Kira} executable, it can also include some arguments for {\Kira}. For example, if the user wants to turn on the parallelization of {\Kira} itself, he or she can set \texttt{KiraCommand} as the following. 
\begin{lstlisting}[language=Mathematica]
KiraCommand = "/some/path/kira --parallel=8";
\end{lstlisting}
This will make {\Kira} run in parallel with maximum job number 8.

The {\Kira} reduction uses {\sc FireFly} by default. Specifically speaking, this means that the interface runs {\Kira} with the following settings in the ``\texttt{jobs.yaml}'' file
\begin{lstlisting}[style=yaml]
run_firefly: true
run_triangular: false
run_back _substitution: false
\end{lstlisting}
If the user wants to use {\sc Fermat} instead, add the following in \texttt{config.txt} 
\begin{lstlisting}[language=Mathematica]
RunFireFlyInKira = False;
\end{lstlisting}
The interface will then run {\Kira} with the following settings in ``\texttt{jobs.yaml}''
\begin{lstlisting}[style=yaml]
run_firefly: false
run_triangular: true
run_back _substitution: true
\end{lstlisting}

Notice that, {\Kira} provide two manners of setting environmental variables for the path of {\sc Fermat}, which is 
\begin{lstlisting}
# sh-shell:
export FERMATPATH="/path/to/Fermat/binary"
# csh-shell:
setenv FERMATPATH "/path/to/Fermat/binary"
\end{lstlisting}
We use \texttt{export} by default. If the user needs to use \texttt{setenv} modify this, he or she can edit the following file in the {\NeatIBP} package path 
\begin{lstlisting}
/package/path/NeatIBP/preload/EnvVarSetter.txt
\end{lstlisting}

The user may want to run the {\Kira} reduction individually. Suppose \texttt{outputPath} is an output path of a finished {\NeatIBP} computation, with complete results (i.e. \texttt{IBP\_all.txt} and so on), the user can use
\begin{lstlisting}[language=bash]
/package/path/NeatIBP/interfaces/Kira/interface/run_kira_reduction.sh -f "/some/path/kira" outputPath
\end{lstlisting}
where \texttt{"-f"} means to force remove the existing temporary or result sub folders in \texttt{outputPath/KiraIO}, including \texttt{firefly_saves/}, \texttt{results/}, \texttt{tmp/} and \texttt{sectormappings/}. 

We can also introduce some arguments for {\Kira} here, as we stated above. Please remember that they must be quoted in command line, like the following. 

\begin{lstlisting}[language=bash]
/package/path/NeatIBP/interfaces/Kira/interface/run_kira_reduction.sh -f "/some/path/kira --parallel=8" outputPath
\end{lstlisting}

In this interface, we also include a tool to translate {\Kira} input files to {\NeatIBP} input files, for the user's convenience. This tool is used individually and is not integrated in the main workflow of {\NeatIBP}. It can be used by
\begin{lstlisting}[language=bash]
math -script KiraToNeatIBP.wl \
--kinematics_file=... \
--integral_families_file=... \
--integral_family_name=... \
--targets_file=... 
\end{lstlisting}
It will by default produce the {\NeatIBP} input files in the directory where the user runs this script.

\subsection{How to use the spanning cuts method in {\NeatIBP}}
\subsubsection{Generating IBP systems on spanning cuts}
In order to use the spanning cuts method in {\NeatIBP}, the user should include the following settings to his or her \texttt{config.txt} of the {\NeatIBP} input files:
\begin{lstlisting}[language=Mathematica]
SpanningCutsMode = True;
\end{lstlisting}

Beware that, the cut naturally breaks the symmetry. As a consequence, in the current version of {\NeatIBP}, we do not yet support symmetry relation when there is any cuts. Thus, in the spanning cuts mode, the user should set the following.
\begin{lstlisting}[language=Mathematica]
NeedSymmetry = False;
\end{lstlisting}

Notice that for some users using {\NeatIBP} with versions earlier than \version, when the spanning cuts method is still under development, they may have used the following setting to turn on the spanning cuts mode:
\begin{lstlisting}[language=Mathematica]
CutIndices = "spanning cuts";
\end{lstlisting}
In the new version, this old ``grammar'' is still accepted but not recommended. It is now interpreted as the following by the new version of {\NeatIBP}:

\begin{lstlisting}[language=Mathematica]
SpanningCutsMode = True;
CutIndices = {};
\end{lstlisting}

The benefit of the new ``grammar'' is that the user can use the spanning cuts method with some indices always on cut. The setting is as follows.
\begin{lstlisting}[language=Mathematica]
SpanningCutsMode = True;
CutIndices = { 1, 2, ... };
\end{lstlisting}
After setting these, {\NeatIBP} builds the sector tree in the cut specified by \texttt{CutIndices}, and then determines the spanning cuts using the strategy introduced in Section \ref{sec:spc}. For example, if a family has bottom sectors as $\{\{1,4,6\},\{1,4,7\},\{2,5,7\},\{2,4,6\},\{3,4,7\}\}$, there are multiple propagators appearing at the 7th position, and the user specifies $\texttt{CutIndices}=\{4\}$, then the spanning cuts will be $\{\{1,4\},\{2,4,6\},\{3,4\}\}$.

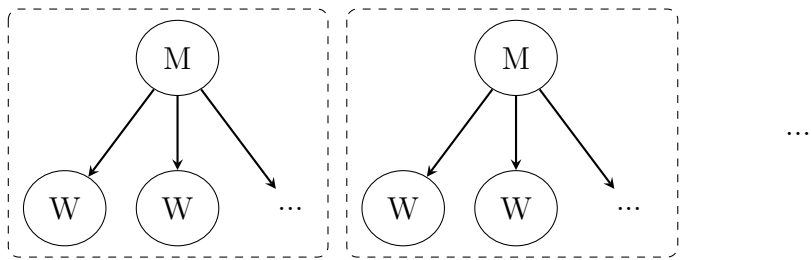
\begin{figure}[htb]
\centering
\begin{tikzpicture}[node distance=3cm,scale=0.5]
  \draw[rounded corners, dashed, draw=black] (-13.5cm,-2.7cm) rectangle (-5cm,-9.3cm);
   \draw[rounded corners, dashed, draw=black] (-4.5cm,-2.7cm) rectangle (4.3cm,-9.3cm);
\node (M1) [item2,yshift=-2cm,xshift=-4.5cm ] {M};
\node (M2) [item2,yshift=-2cm,xshift=0cm ] {M};
\node (Mn) [plaintext,yshift=-3cm,xshift=3.75cm ] {...};
\node (W11) [item2,yshift=-4cm,xshift=-6cm ] {W};
\node (W12) [item2,yshift=-4cm,xshift=-4.5cm ] {W};
\node (W1n) [plaintext,yshift=-4cm,xshift=-3cm ] {...};
\node (W21) [item2,yshift=-4cm,xshift=-1.5cm ] {W};
\node (W22) [item2,yshift=-4cm,xshift=0cm ] {W};
\node (W2n) [plaintext,yshift=-4cm,xshift=1.5cm ] {...};
\draw [arrow] (M1) -- (W11);
\draw [arrow] (M1) -- (W12);
\draw [arrow] (M1) -- (W1n);
\draw [arrow] (M2) -- (W21);
\draw [arrow] (M2) -- (W22);
\draw [arrow] (M2) -- (W2n);

\end{tikzpicture}
\caption{The two-layer parallelization in spanning cuts mode of {\NeatIBP}. The first layer is parallelization between managers. The second layer is parallelization between workers that belong to a manager. In this figure, ``M'' stands for a manager and ``W'' stands for a worker.}
\label{fig: M-W}
\end{figure}

After finding the spanning cuts, {\NeatIBP} launches individual sub jobs to generate IBP identities on different cuts. The working folders of these sub jobs are set as
\begin{lstlisting}
your_outputPath/tmp/spanning_cuts_missions/cut_*/
\end{lstlisting}
The computation on of each sub job employs a manager-worker framework as introduced in Fig. 2 of Ref. \cite{Wu:2023upw}. These sub jobs are by default running in parallel as shown in Fig. \ref{fig: M-W}. If the user wants to run them in a sequential manner, set in the \texttt{config.txt} with
\begin{lstlisting}[language=Mathematica]
SpanningCutsEvaluationMode = "Sequential";
\end{lstlisting}

The user may wish to set up a limit for the usage of parallelization. Practically, {\NeatIBP} v1.1 allows user to set up a limit for the total number of {\sc{Mathematica}} kernel used in the whole computation. To do so, the user needs to include the following setting in his or her \texttt{config.txt}:
\begin{lstlisting}[language=Mathematica]
MathKernelLimit = n;
\end{lstlisting}
where $n$ stands for the limit of the number of {\sc{Mathematica}} kernels the user wants to set. The default value of this setting is \texttt{Infinity}, which means setting no limit on this usage.

\begin{figure}[htb]
\centering
\begin{tikzpicture}[node distance=3cm,scale=0.5]
  \draw[rounded corners, dashed, draw=black] (-13.5cm,-2.7cm) rectangle (-5cm,-9.3cm);
   \draw[rounded corners, dashed, draw=black] (-4.5cm,-2.7cm) rectangle (4.3cm,-9.3cm);
\node (HQ) [item2,text width=1cm] {HQ};
\node (M1) [item2,yshift=-2cm,xshift=-4.5cm ] {M};
\node (M2) [item2,yshift=-2cm,xshift=0cm ] {M};
\node (Mn) [plaintext,yshift=-2cm,xshift=4.5cm ] {...};
\node (W11) [item2,yshift=-4cm,xshift=-6cm ] {W};
\node (W12) [item2,yshift=-4cm,xshift=-4.5cm ] {W};
\node (W1n) [plaintext,yshift=-4cm,xshift=-3cm ] {...};
\node (W21) [item2,yshift=-4cm,xshift=-1.5cm ] {W};
\node (W22) [item2,yshift=-4cm,xshift=0cm ] {W};
\node (W2n) [plaintext,yshift=-4cm,xshift=1.5cm ] {...};
\draw [arrow] (HQ) -- (M1);
\draw [arrow] (HQ) -- (M2);
\draw [arrow] (HQ) -- (Mn);
\draw [arrow] (M1) -- (W11);
\draw [arrow] (M1) -- (W12);
\draw [arrow] (M1) -- (W1n);
\draw [arrow] (M2) -- (W21);
\draw [arrow] (M2) -- (W22);
\draw [arrow] (M2) -- (W2n);

\end{tikzpicture}
\caption{The HQ-manager-worker framework in {\NeatIBP} when \texttt{MathKernelLimit} is set to be a finite integer, where ``M'' stands for a manager and ``W'' stands for a worker.}
\label{fig: HQ-M-W}
\end{figure}
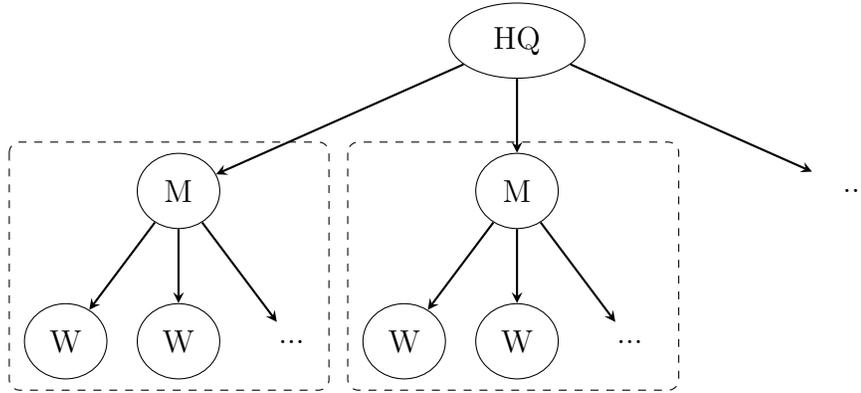

When \texttt{MathKernelLimit} is set to be an finite integer number, {\NeatIBP} will launch an ``HQ'' process to distribute kernels to the managers. This framework is shown in Fig. \ref{fig: HQ-M-W}. In this framework, each activated node, including HQ, manager, or worker, occupies one {\sc{Mathematica}} kernel. In the current version, the strategy of distribution is that, HQ activates as many managers as allowed. The allowed number of managers activating simultaneously, say, $n$, can be set in \texttt{config.txt} by
\begin{lstlisting}[language=Mathematica]
MaxManagers = n;
\end{lstlisting}

In the current version, the default value of this setting is 4. For efficiency reason, the user may need to tune this setting by themselves. On the one hand, \texttt{MaxMagagers} should not be too large, in case that too many kernels are  occupied by managers. Then there will be few activated workers that really perform the computations. On the other hand, \texttt{MaxMagagers} should not be too small, in case that too many managers, as well as their workers, are waiting for activation. One extremal case of this is setting \texttt{MaxMagagers} as 1. In this case, it is equivalent to a computation sequential between the managers.

We remind that, \textit{launching a monitor using }\texttt{monitor.sh} \textit{ occupies an additional {\sc Mathematica} kernel,} as stated in \cite{Wu:2023upw}. The user is recommended to reserve a kernel for monitor. 

The way of running a monitor is the same as that introduced in \cite{Wu:2023upw}. In spanning cuts mode, the monitor shows the progress of all the cuts. The following is an example.

\begin{lstlisting}[breaklines=false,columns=fullflexible,basicstyle=\fontsize{4pt}{8pt}\ttfamily]
----------------------------------------------
2025.1.10 18:27:1
----------------------------------------------
cut         	status     	progress	waiting	computable	computing	finished	*lost	*unlabelled
{1, 3, 5, 7}	finished   	100.%   	0      	0         	0        	16      	0    	0          
{1, 4, 5, 7}	finished   	100.%   	0      	0         	0        	16      	0    	0          
{1, 4, 6, 8}	finished   	100.%   	0      	0         	0        	16      	0    	0          
{1, 5, 8}   	in progress	18.8%   	26     	0         	0        	6       	0    	0          
{2, 4, 5, 7}	finished   	100.%   	0      	0         	0        	16      	0    	0          
{2, 5, 8}   	in progress	71.9%   	5      	0         	4        	23      	0    	0          
{2, 6, 8}   	in progress	50.%    	11     	0         	5        	16      	0    	0          
{3, 6, 8}   	in progress	50.%    	11     	0         	5        	16      	0    	0          
{3, 7, 8}   	in progress	78.1%   	4      	0         	3        	25      	0    	0          
{4, 7, 8}   	in progress	78.1%   	4      	0         	3        	25      	0    	0   
\end{lstlisting}
The numbers in the table displayed by the monitor are the number of sectors of the corresponding status.

After running {\NeatIBP} in spanning cuts mode, the results are in the sub folders as
\begin{lstlisting}
your_outputPath/results/results_spanning_cuts/cut_*/
\end{lstlisting}
where, like the case in normal mode (see \cite{Wu:2023upw}), \texttt{IBP\_all.txt} is the system of IBP indentities generated on the corresponding cut, and \texttt{MI\_all.txt} the master integrals, \texttt{OrderedIntegrals.txt} the relevant integrals ordered by preference.  

By default, the spanning cuts mode generates the equations only. To continue the following processes shown in Fig. \ref{fig: work flow spc}, one have to turn on \texttt{PerformIBPReduction} (which is by default \texttt{False}). And, one also has to specify the \texttt{IBPReductionMethod} (currently the only practical choice is \texttt{"Kira"}). The corresponding settings are what we introduced above:
\begin{lstlisting}[language=Mathematica]
PerformIBPReduction = True;
IBPReductionMethod = "Kira";
KiraCommand = "/some/path/kira";
FermatPath = "/some/path/fer64";
\end{lstlisting}

After setting these, {\NeatIBP} runs the remaining procedures in Fig. \ref{fig: work flow spc} automatically. The final result can be found at

\begin{lstlisting}
your_outputPath/results/reduced_IBP_spanning_cuts_merged/from_Kira_reduction/IBPTable.txt
\end{lstlisting}

In the next several sub sub sections, we introduce some advanced settings concerning the sub steps including the consistency check, IBP shortening, IBP reduction, and merging.

\subsubsection{The consistency check}\label{sec: consistency check manual sub sub section}
As stated in Section \ref{sec:spc}, the consistency check is in principle a mandatory step. Though, it can be turned off by 
\begin{lstlisting}[language=Mathematica]
SpanningCutsConsistencyCheck = False;
\end{lstlisting}
However, it is not recommended to do so. Also, when the consistency check is turned off, the user is not allowed to use the ``Shorten IBP'' module.

The consistency check step in Fig. \ref{fig: work flow spc} is by default running in parallel. If the user wants to run them sequentially, add the following in \texttt{config.txt} 
\begin{lstlisting}[language=Mathematica]
ConsistencyCheckParallelization = False;
\end{lstlisting}

The parallelization by default sets jobs once for all in parallel. To limit the number of jobs that run simultaneously, set in \texttt{config.txt}
\begin{lstlisting}[language=Mathematica]
ConsistencyCheckParallelJobNumber = n;
\end{lstlisting}
where $n$ is the maximum number of jobs (the default value for this setting is \texttt{Infinity}). If one sets the job number to a finite number, the parallel jobs are arranged in a naive way. They are divided into different groups and run sequentially in parallel group by group. For example, if one sets the max job number as 4, {\NeatIBP} runs jobs number 1 to 4 in parallel first. Only when all four jobs are finished, {\NeatIBP} starts to run jobs number 5 to 8. This manner is indeed not the most efficient way. We may expect that if one of the jobs among numbers 1 to 4 finished, the job number 5 should immediately start. We can achieve this by using \texttt{GNU parallel} \cite{Tange2011a,tange2018gnu}. To do this, one should already have \texttt{GNU parallel} installed on his or her device following \url{https://www.gnu.org/software/parallel/} or \cite{Tange2011a}, and then set
\begin{lstlisting}[language=Mathematica]
UseGNUParallel = True;
ConsistencyCheckParallelizationMethod = "GNUParallel";
\end{lstlisting}
The default value of the second setting above is \texttt{"Naive"}. Please note that the authors of \texttt{GNU parallel} request to be cited. Thus, there is a setting \texttt{UseGNUParallel} which is the main switch to turn on all modules in {\NeatIBP} that relies on \texttt{GNU parallel}. This main switch also serves as a reminder, if you set \texttt{UseGNUParallel = True} when using {\NeatIBP} in your work, please cite \texttt{GNU parallel} \cite{Tange2011a,tange2018gnu} as well.

By default, the command that runs \texttt{GNU parallel} is ``\texttt{parallel}'' on most of the devices. If this is not the case on your device, please set the following in your \texttt{config.txt}:
\begin{lstlisting}[language=Mathematica]
GNUParallelCommand = "your GNU parallel command";
\end{lstlisting}

The user may want to run the consistency check individually. Suppose \texttt{outputPath} is an output path of a finished {\NeatIBP} computation, and the {\sc Mathematica} command is \texttt{math}, the user can use the following command to do so
\begin{lstlisting}[language=bash]
math -script /package/path/NeatIBP/FFSpanningCutsConsistencyCheck.wl outputPath
\end{lstlisting}

\subsubsection{Shortening the IBPs on cut}
The IBP shortening step in Fig. \ref{fig: work flow spc} is an optional step, though it is included by default. If the user wants to skip this step, set in \texttt{config.txt} the following:
\begin{lstlisting}[language=Mathematica]
ShortenSpanningCutsIBPs = False;
\end{lstlisting}
After setting this, the reduction will be applied to the IBP system without shortening.

If the user wants to shorten the IBP system, but does not want to use them in the next reduction steps, use the following setting instead
\begin{lstlisting}[language=Mathematica]
UseShortenedSpanningCutsIBPs = False;
\end{lstlisting}

The user may want to run the IBP shortening individually. Suppose \texttt{outputPath} is an output path of a finished {\NeatIBP} computation, and the {\sc Mathematica} command is \texttt{math}, the user can use the following command to do so
\begin{lstlisting}[language=bash]
math -script /package/path/NeatIBP/SpanningCutsIBPShorten.wl outputPath
\end{lstlisting}

\subsubsection{Reducing the IBP systems on each cut}
The IBP reduction steps consist of multiple individual reduction processes as introduced in Section \ref{sec: kira interface manual}. These reduction jobs are by default in parallel. To run them in a sequential way, set the following in \texttt{config.txt}
\begin{lstlisting}[language=Mathematica]
SPCIBPReductionParallelization = False;
\end{lstlisting}

Also here, as introduced in the Section \ref{sec: consistency check manual sub sub section}, one can set a limit $n$ for the paralleled reduction jobs using 
\begin{lstlisting}[language=Mathematica]
SPCIBPReductionParallelJobNumber = n;
\end{lstlisting}
The default value of this setting is \texttt{Infinity}. Also, like that in Section \ref{sec: consistency check manual sub sub section}, the parallelization strategy is \texttt{"Naive"} by default. User can use \texttt{GNU parallel} by setting the following in \texttt{config.txt}
\begin{lstlisting}[language=Mathematica]
UseGNUParallel = True;
SPCIBPReductionParallelizationMethod = "GNUParallel";
\end{lstlisting}
Still, please cite \texttt{GNU parallel} \cite{Tange2011a,tange2018gnu} if you use it.

\subsubsection{Merging reduced IBPs on spanning cuts}

After the reduction on each cut, {\NeatIBP} will automatically merge them using the strategy introduced in Section \ref{sec:spc}. 

If the user wants to run this step by hand, suppose \texttt{outputPath} is an output path of a finished {\NeatIBP} computation, and the {\sc Mathematica} command is \texttt{math}, he or she can use the following commands
\begin{lstlisting}[language=bash]
math -script /package/path/NeatIBP/ReducedSpanningCutsMerge.wl -Kira outputPath
\end{lstlisting}
or
\begin{lstlisting}[language=bash]
math -script /package/path/NeatIBP/ReducedSpanningCutsMerge.wl -s -Kira outputPath
\end{lstlisting}
where the setting ``\texttt{-Kira}'' tells the script it should read in the reduction results produced by {\Kira}. The setting ``\texttt{-s}'' tells the script that the reduced IBP results are from a shortened IBP system (if not, run without ``\texttt{-s}''). Notice that, an incorrect setting of ``\texttt{-s}'' will lead to a wrong result. Currently, {\NeatIBP} does not provide any check on this, the users should be careful by themselves. If the user is not sure about this, he or she can refer to the last line of the script file 
\begin{lstlisting}
outputPath/tmp/assigned_reduction_script.sh
\end{lstlisting}
for the correct setting, which is the merging command that were run in the automated work flow.

\subsection{Using the syzygy simplification algorithms}\label{sec: syzygy simplification manual}

In this section we introduce some settings relevant to the syzygy simplification algorithms introduced in \ref{sec: syzygy simplification algorithm}. First of all, to turn on this method, use the following setting in \texttt{config.txt}
\begin{lstlisting}[language=Mathematica]
SimplifySyzygyVectorsByCut = True;
\end{lstlisting}
With this setting, {\NeatIBP} runs Algorithm \ref{algorithm: lift selection} first and then runs Algorithm \ref{algorithm: further selection}. If the user wants to skip the Algorithm \ref{algorithm: lift selection}, use the following setting in \texttt{config.txt}
\begin{lstlisting}[language=Mathematica]
SkipLiftSelection = True;
\end{lstlisting}
If the user wants to skip the Algorithm \ref{algorithm: further selection}, use the following setting in \texttt{config.txt}
\begin{lstlisting}[language=Mathematica]
FurtherSyzygyVectorsSelection = False;
\end{lstlisting}

As stated in Section \ref{sec: syzygy simplification algorithm}, there are ``strictness'' setting to randomly reserve some of the generators that should be deleted. The corresponding settings are 
\begin{lstlisting}[language=Mathematica]
LiftSelectionStricty = 0.xxx;
FurtherSyzygyVectorsSelectionStricty = 0.xxx;
\end{lstlisting}
for the strictness of Algorithm \ref{algorithm: lift selection} and \ref{algorithm: further selection}, respectively. Notice that the values should be between 0 and 1. A value greater than 1 will be interpreted as 1, while a negative value will be interpreted as 0.

Also, as stated in Section \ref{sec: syzygy simplification algorithm}, the Gr{\"o}bner basis computations in these algorithms are performed numerically on finite fields. The numerical point is chosen as \texttt{GenericPoint} and \texttt{GenericD}, and the finite field modulus can be set as
\begin{lstlisting}[language=Mathematica]
FFiniteFieldModulus2 = 117763;
\end{lstlisting}
which is specially for these syzygy simplification algorithms, where $117763$ is its default value. Since this value will not be used in {\sc SpaSM}, it can be larger than $46337$.

Notice that, although the Gr{\"o}bner basis computations in these algorithms are performed numerically, for some complicated examples, it could still take a long time. As stated in Section \ref{sec:syzygy simplification comments}, the user may want to introduce some time constraints. For Algorithm \ref{algorithm: lift selection}, the user can use the following setting in \texttt{config.txt} to limit the {\Singular} running time. 
\begin{lstlisting}[language=Mathematica]
LiftSelectionSingularTimeConstraint = t;
\end{lstlisting}
When the {\Singular} computation in this algorithm, including the \Gr basis computation and the lift computation, exceeds $t$ seconds, {\Singular} terminates, and the program gives up the simplification via Algorithm \ref{algorithm: lift selection}.

The user can also set some limit on the time used in Algorithm \ref{algorithm: further selection}. Specifically, set
\begin{lstlisting}[language=Mathematica]
FurtherSelectionSingularTimeConstraint = t;
\end{lstlisting}
Then, if a \texttt{Singular} process for the \Gr basis computation in Algorithm \ref{algorithm: further selection} takes a time longer than $t$ seconds, \texttt{Singular} stops and the corresponding generator will be kept. There is another setting as
\begin{lstlisting}[language=Mathematica]
FurtherSelectionTimeUsedLimit = t;
\end{lstlisting}
restricting the total time used in the selection step (i.e. the \texttt{for} loop) in Algorithm \ref{algorithm: further selection}. If the time used in Algorithm \ref{algorithm: further selection} exceeds $t$ seconds, it stops and keeps the rest generators that are not yet scanned in it. The user is not recommended to set \texttt{FurtherSelectionTimeUsedLimit} only without setting \texttt{FurtherSelectionSingularTimeConstraint}. Because the former setting checks the total time used only after {\Singular} finishes. If a {\Singular} process runs for too-long a time, without the later setting, the program is not able to detect that the total time is exceeding the former constraint.

In Section \ref{sec:syzygy simplification comments}, we have stated that there is a way to not use algorithms in Section \ref{sec: mc simp no mp}, even if the target integrals are all without multiple propagators. The corresponding setting in \texttt{config.txt} is\footnote{D(enominator)-corner integrals in the setting stands for integrals without multiple propagators.}
\begin{lstlisting}[language=Mathematica]
AllowingDCornerOnlyModeInSimplifyByCut = False;
\end{lstlisting}

Also, in Section \ref{sec:syzygy simplification comments}, we stated that there is an older method named \texttt{NCornerKill} that can be turned off. The corresponding setting in \texttt{config.txt} is
\begin{lstlisting}[language=Mathematica]
FIBPsNCornerKill = False;
\end{lstlisting}

\subsection{Other instructions}
In this sub section, we include some useful instructions that is updated since the version introduced in \cite{Wu:2023upw}.

\subsubsection{{\Singular} related instructions}\label{sec:singular related instructions}
In Section \ref{sec: other feature}, we introduced some {\Singular} related new features. In this sub sub section, we give the corresponding instructions.

As stated in Section \ref{sec: other feature}, the user can use the flexible degree bound method in module intersection computation. The corresponding settings in \texttt{config.txt} are

\begin{lstlisting}[language=Mathematica]
FlexibleNeatIBPIntersectionDegreeBound = True;
NeatIBPIntersectionTimeConstraintForFlexibleDegreeBound = t;
NeatIBPIntersectionDegreeBoundDecreaseLimit = n;
\end{lstlisting}
where $t$ is the number of seconds limited for {\Singular}, with default value as $3600$, and $n$ is the degree-decrease limit with default value as 2. With these default settings, suppose the initial degree bound is $6$, {\NeatIBP} tries to run {\Singular} to solve the module intersection problem with degree bound as $6$. If the time used exceeds $3600$ seconds, it tries again with degree bound as $5$. If the time still exceeds, then it redo with degree bound as $4$. Finally, if time exceeds again, it returns ``\texttt{\$Failed}''.

In Section \ref{sec: other feature}, we also introduced that we can set the monomial orderings for {\Singular}. The orderings can be referred to in 
\begin{center}
\url{https://www.singular.uni-kl.de/Manual/4-4/sing_985.htm}
\end{center}

In {\NeatIBP}, the Baikov variables (i.e., the propagators) and the kinematic parameters are all treated as ring variables in {\Singular}. They are categorized into three groups: propagators in the denominator, the propagators in the numerator (ISP), and the kinematic variables. For convenience of description, we take the massless two-loop five-point diagram as an example. It has 8 propagators in the denominator $z_1\sim z_8$, 3 ISP $z_9\sim z_{11}$ and 5 parameters $s_{12},s_{23},s_{34},s_{45},s_{15}$. By default, the definition of the ring in {\Singular} will be
\begin{lstlisting}
ring r=0,(z9,z10,z11,z1,...,z8,c1,...,c5),(dp(11),dp(5));
\end{lstlisting}
In the above, \texttt{c1},...,\texttt{c5} stands for the parameters $s_{12},\cdots,s_{15}$. The ordering is a block ordering putting the Baikov variables and parameters into different blocks, applying the degree reverse lexicographical ordering (\texttt{dp}) in the blocks. And, in the Baikov variable block, the numerators (ISP) are in front of the denominators. 

To change the default ordering, the user can use the following settings. Firstly, the user can change the ordering \texttt{dp} to \texttt{Dp}, \texttt{lp} or \texttt{rp}. For example, after setting the following in \texttt{config.txt}
\begin{lstlisting}[language=Mathematica]
SingularMonomialOrdering = "Dp";
\end{lstlisting}
the {\Singular} ordering of the above example will be 
\begin{lstlisting}
ring r=0,(z9,z10,z11,z1,...,z8,c1,...,c5),(Dp(11),Dp(5));
\end{lstlisting}

The ordering also supports the weighted orderings \texttt{wp} and \texttt{Wp}. Although the two orderings have high freedom when used in {\Singular}, in the current version of {\NeatIBP}, we use them limitedly to ignore the parameters when counting the degrees, for the cases where the degree bound is set. Specifically, for example, if the user sets the following in \texttt{config.txt}
\begin{lstlisting}[language=Mathematica]
SingularMonomialOrdering = "wp";
\end{lstlisting}
the {\Singular} ordering of the above example will be 
\begin{lstlisting}
ring r=0,(z9,z10,z11,z1,...,z8,c1,...,c5),(wp(1,1,1,1,1,1,1,1,1,1,1),wp(0,0,0,0,0));
\end{lstlisting}

The user can put the denominators in front of the numerators. To do so, set the following in \texttt{config.txt}
\begin{lstlisting}[language=Mathematica]
SingularBaikovVariableOrdering = "DenominatorFirst";
\end{lstlisting}
Then, the {\Singular} ordering of the above example will be 
\begin{lstlisting}
ring r=0,(z1,...,z8,z9,z10,z11,c1,...,c5),(dp(11),dp(5));
\end{lstlisting}

The user can divide the Baikov variables into two blocks in the block ordering, categorized by numerator and denominator. To do so, set the following in \texttt{config.txt}
\begin{lstlisting}[language=Mathematica]
SingularBaikovVariableBlockOrdering = True;
\end{lstlisting}
Then, the {\Singular} ordering of the above example will be 
\begin{lstlisting}[breaklines=false,columns=fullflexible]
ring r=0,(z9,z10,z11,z1,...,z8,c1,...,c5),(dp(3),dp(8),dp(5));
\end{lstlisting}

The above settings can be used together. For example, the user can set the following.
\begin{lstlisting}[language=Mathematica]
SingularMonomialOrdering = "wp";
SingularBaikovVariableOrdering = "DenominatorFirst";
SingularBaikovVariableBlockOrdering = True;
\end{lstlisting}
Then, the {\Singular} ordering of the above example will be 
\begin{lstlisting}
ring r=0,(z1,...,z8,z9,z10,z11,c1,...,c5),(wp(1,1,1,1,1,1,1,1),wp(1,1,1),wp(0,0,0,0,0));
\end{lstlisting}

The user may want to permute the ordering of the Baikov variables or the parameters. To permute the Baikov variables, the user has to do that in the \texttt{kinematicsFile} in the current version. To permute the parameters, using the above example, he or she can set
\begin{lstlisting}[language=Mathematica]
ParameterRepermute = True;
ParameterRepermuteIndex = {2,3,5,1,4};
\end{lstlisting}
This changes the correspondence between \texttt{c1},...,\texttt{c5} and $s_{12},\cdots,s_{15}$. Specifically, after setting the above, \texttt{c1},...,\texttt{c5} stands for $s_{23},s_{34},s_{15},s_{12},s_{45}$ respectively.

\subsubsection{Instructions about the symmetry algorithms}\label{sec: symmetry manual}
As stated above, the new version of {\NeatIBP} has implemented the symmetry algorithms introduced in \cite{Wu:2024paw}. We list some related settings in this sub sub section. 

As stated in \cite{Wu:2024paw}, there are two options for external momenta transformation algorithms, and two options for external momenta grouping algorithms. In \texttt{config.txt}, the user can use 
\begin{lstlisting}[language=Mathematica]
PreferedExternalExtendedRotationMethod = "DeltaPlaneProjection";
\end{lstlisting}
or
\begin{lstlisting}[language=Mathematica]
PreferedExternalExtendedRotationMethod = "Orthogonalization";
\end{lstlisting}
to choose the preferred transformation algorithm. If the preferred algorithm encounters the vanishing-Gram-determinant problem stated in \cite{Wu:2024paw}, {\NeatIBP} automatically switches to another one. The default choice of this setting is ``\texttt{DeltaPlaneProjection}''.

The user can use the settings
\begin{lstlisting}[language=Mathematica]
ExternalMomentaGroupingMethod = "MomentumSpace";
\end{lstlisting}
or
\begin{lstlisting}[language=Mathematica]
ExternalMomentaGroupingMethod = "FeynmanParameterization";
\end{lstlisting}
to set the grouping algorithm. Currently, the default value for this setting is ``\texttt{MomentumSpace}''.

In the symmetry algorithms, there are steps in solving algebraic equations of the undetermined coefficients from the momentum transformation ansatz. This is usually not time-consuming, but there could be some cases in which it is. There is a setting of time limit on this step. If the time used in solving the equations exceeds the limit, {\NeatIBP} gives up the solution for the current symmetry transformation and considers the corresponding symmetry is not found. The corresponding setting is
\begin{lstlisting}[language=Mathematica]
MomentumMapTimeConstraint = t;
\end{lstlisting}
where $t$ is the time limit with the unit as seconds. The default value of this setting is 15 in the current version. 

\subsubsection{To set {\sc Mathematica} command and shell processor}
In the earlier versions of {\NeatIBP}, the {\sc Mathematica} command it uses is hard coded as \texttt{"math"}. In the latest versions, it can be specified by editting the following file in the {\NeatIBP} package path 
\begin{lstlisting}
/package/path/NeatIBP/preload/MathematicaCommand.txt
\end{lstlisting}
Similarly, the shell processor can also be specified by\footnote{But we do not recommend the user modify this if not necessary, since the other shell processors are not well tested.} 
\begin{lstlisting}
/package/path/NeatIBP/preload/ShellProcessor.txt
\end{lstlisting}

\subsubsection{To run {\NeatIBP} in a different folder}
In the earlier versions of {\NeatIBP}, the user must run it in the same folder of the input files including \texttt{config.txt}. In the new version, one can run it from another folder, by adding the file name of \texttt{config.txt} as an argument, as follows
\begin{lstlisting}[language=bash]
/package/path/NeatIBP/run.sh /working/path/config.txt
\end{lstlisting}
This is similar for \texttt{continue.sh} and \texttt{monitor.sh}.

\subsubsection{To kill {\NeatIBP} processes}
When one runs {\NeatIBP}, it will create several sub processes. Sometimes, a simple ``Ctrl+C'' in the terminal will not kill all the sub processes. They could instead become orphan processes\footnote{We leave the improvement for this problem to the updates towards the future versions.}. The user might want to search for the process ids for all these orphan processes in order to kill them manually. A good way to do so is to search by the name of \texttt{outputPath}. For example, if one {\NeatIBP} mission runs with \texttt{outputPath} as
\begin{lstlisting}[language=Mathematica]
"/some/path/outputs/somename/"
\end{lstlisting}
Then, run
\begin{lstlisting}[language=Mathematica]
ps -ef | grep /some/path/outputs/somename/
\end{lstlisting}
This will find most of the sub processes that are related to the output path, as well as their process ids. Using linux shell tools, for example, \texttt{awk} \cite{robbins2015effective}, the user can kill them in an efficient way.

\section{Examples}\label{sec: examples}
In this section we give some examples to show the performance of the new version of {\NeatIBP}. The input files will be updated into
\begin{lstlisting}
examples/examples_in_the_papers/[the arXiv number of this paper]/[the example name]
\end{lstlisting}
In the input files, we omitted some unimportant settings including the dependency paths, resource used limit, etc. The examples in this section, if not specified, were performed on a machine with 20 CPU threads and 120 GB memory. The {\NeatIBP} version used to run these examples is {\versionexample}.

\subsection{A massless two-loop five-point example}
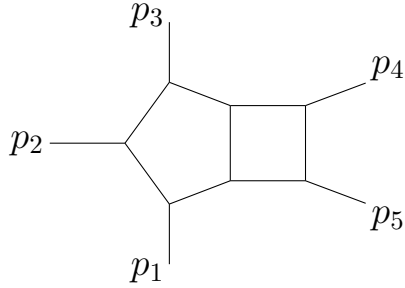
\begin{figure}[hbtp]
\centering
\begin{tikzpicture}[scale=1]
\draw (1.4,-0.5) -- (0.59,-0.81);
\draw (0,0) -- (0.59,-0.81);
\draw(0,0) -- (0.59,0.81);
\draw(1.4,0.5) -- (0.59,0.81);
\draw (1.4,0.5) -- (2.4,0.5);
\draw(2.4,0.5) -- (2.4,-0.5);
\draw (1.4,-0.5) -- (2.4,-0.5);
\draw (1.4,0.5) -- (1.4,-0.5);

\draw (0.59,-1.61) -- (0.59,-0.81);
\draw(-1,0)--(0,0);
\draw (0.59,1.61) -- (0.59,0.81);
\draw(2.4,0.5)--(3.2,0.8);
\draw(2.4,-0.5)--(3.2,-0.8);

\node[font=\large\bfseries] at (0.3,-1.7){$p_1$};
\node[font=\large\bfseries] at (-1.3,0){$p_2$};
\node[font=\large\bfseries] at (0.3,1.7){$p_3$};
\node[font=\large\bfseries] at (3.5,1){$p_4$};
\node[font=\large\bfseries] at (3.5,-1){$p_5$};

\end{tikzpicture}
\caption{The massless pentagon-box diagram}
\label{fig:pentabox}
\end{figure}

In this section, we give a two-loop five-point example to show the usage of the {\Kira} interface. The diagram is shown in Fig. \ref{fig:pentabox}. This example can be found in the example folder stated above with the example name ``\texttt{pentabox}''. The propagators are
\begin{equation}
    \begin{aligned}
        &D_1=l_1^2,\quad
        D_2=(l_1+p_1)^2,\quad
        D_3=(l_1+p_1+p_2)^2,\\
        &D_4=(l_1+p_1+p_2+p_3)^2,\quad
        D_5=(l_2-p_1-p_2-p_3)^2,\\
        &D_6=(l_2+p_5)^2,\quad
        D_7=l_2^2,\quad
        D_8=(l_1+l_2)^2,\\
        &D_9=(l_2+p_1)^2,\quad
        D_{10}=(l_2+p_1+p_2)^2,\\
        &D_{11}=(l_1+p_5)^2.
    \end{aligned}
\end{equation}
with $D_9\sim D_{11}$ irreducible scalar products (ISP). The kinematics conditions are
\begin{equation}
    \begin{aligned}
        &p_1^2=p_2^2=p_3^2=p_4^2=p_5^2=0,\\
        &(p_1+p_2)^2=s_{12},\quad
        (p_2+p_3)^2=s_{23},\quad
        (p_3+p_4)^2=s_{34},\\
        &(p_4+p_5)^2=s_{45},\quad
        (p_1+p_5)^2=s_{15},\quad
    \end{aligned}
\end{equation}
where $p_1+p_2+p_3+p_4+p_5=0$.

For the purpose of demonstration of the interface, we choose the target integrals as integrals from the top sector, without multiple propagators, and with the numerator degree as 4 or 5. We turned off symmetry relations because we needed to compare the results with and without using the spanning cuts method.

We consider targets with numerator degree 4 as a lighter example first. We run this example in the normal mode, i.e. without spanning cuts. In the normal mode, {\NeatIBP} used 18 minutes to generate 3708 IBP identities, taking $3.0$ MB of disk space. The number of master integrals is 62. The conversion of IBP identities to {\Kira} format took less than 1 minute. The {\Kira} reduction, took 24 hours. The resulting reduced IBP table took 43.6 MB of disk space. In the {\Kira} reduction, we used {\FireFly}. Notice that, in this example, we did not use the parallelization in {\Kira}, i.e., setting ``\texttt{--parallel=...}'', in order to give a clearer estimation of the CPU resources needed in the reduction.

We then run this example in the spanning cuts mode by adding 
\begin{lstlisting}[language=Mathematica]
SpanningCutsMode = True;
\end{lstlisting}
in the ``\texttt{config.txt}'' file. While running in the spanning cuts mode, {\NeatIBP} found 10 spanning cuts. Their performance is shown in Tab. \ref{tab: pentabox example spc}.
\begin{table}[h]
    \centering
    \begin{tabular}{|c|c|c|c|c|c|c|}
    \hline
         cut & \#MI&\#IBP&$T_{\text{gen}}$&IBP size&$T_{\text{red}}$ & reduced IBP size\\
         \hline
    
         $\{1,3,5,7\}$ & $ 13 $ &$ 333 $& 3m18s & 172K & 2m & 590K\\
         \hline
         $\{1,4,5,7\}$ & $ 9 $ &$ 310 $& 3m21s & 132K & $<$1m & 251K\\
         \hline
         $\{1,4,6,8\}$ & $ 12 $ &$ 474 $& 5m5s & 271K & 1m & 525K\\
         \hline
         $\{1,5,8\}$ & $ 21 $ &$ 781 $& 6m56s & 510K & 1h26m & 4.5M\\
         \hline
         $\{2,4,5,7\}$ & $ 13 $ &$ 315 $& 3m38s & 174K & 6m & 1.1M\\
         \hline
         $\{2,5,8\}$ & $ 27 $ &$ 813 $& 7m47s & 615K & 5h23m & 14M\\
         \hline
         $\{2,6,8\}$ & $ 31 $ &$ 1583 $& 8m12s & 1.2M & 2h53m & 3.9M\\
         \hline
         $\{3,6,8\}$ & $ 31 $ &$ 1372 $& 7m56s & 1.1M & 4h3m & 4.7M\\
         \hline
         $\{3,7,8\}$ & $ 27 $ &$ 1044 $& 7m16s & 706K & 4h53m & 8.6M\\
         \hline
         $\{4,7,8\}$ & $ 21 $ &$ 1045 $& 7m5s & 658K & 1h28m & 4.1M\\
         \hline

    \end{tabular}
    \caption{The performance of all cuts in this pentagon-box example with numerator degree 4. The second and the third column is the number of master integrals and IBP identities, respectively. The fourth column is the {\NeatIBP} time used for generating the IBP system. The fifth column is the disk size of the IBP system generated by {\NeatIBP}. The sixth column is the time used for the {\Kira} reduction including the interface. The seventh column is the disk size of the reduced IBP table. Notice that, for the sixth column, we still did not turn on parallelization of {\Kira}, like what we did in the normal mode.}
    \label{tab: pentabox example spc}
\end{table}

The merging step in this example took 1 minute. The merged result takes 42 MB disk space. The total process of the spanning cuts method took 5 hours and 33 minutes. We have checked the final results from the two different approaches, i.e. with and without spanning cuts mode. They agree with each other.

We also ran this diagram with numerator degree 5. In this situation, the IBP reduction step is relatively much more time consuming than degree 4. Thus, we used ``\texttt{--parallel=20}'' in {\Kira} both in the normal mode and in the spanning cuts mode. Besides, in order to avoid interference between the IBP reduction between spanning cuts and the overuse of memory, we set
\begin{lstlisting}[language=Mathematica]
SPCIBPReductionParallelization = False;
\end{lstlisting}
in the spanning cuts mode.

In the normal mode, {\NeatIBP} used 19 minutes to generate $9488$ IBP identities with $62$ master integrals. The disk size of the IBP identities is $6.9$ MB. The conversion to {\Kira} of the IBP identities took less than 1 minute. The {\Kira} reduction took about 94 hours. Notice that, although this is running with ``\texttt{--parallel=20}'', the CPU time needed is not this value times 20, since some CPU threads were sometimes vacant while running. This phenomenon also appears in the spanning cuts mode. The final result of the reduced IBP table has a disk size of $253.4$ MB.

The performance in the spanning cuts mode is shown in Tab. \ref{tab: pentabox example spc deg5}.

\begin{table}[h]
    \centering
    \begin{tabular}{|c|c|c|c|c|c|c|}
    \hline
         cut & \#MI&\#IBP&$T_{\text{gen}}$&IBP size&$T_{\text{red}}$ & reduced IBP size\\
         \hline
    
         $\{1,3,5,7\}$ & $ 13 $ &$ 697 $& 3m21s & 368K & 3m & 2.5M\\
         \hline
         $\{1,4,5,7\}$ & $ 9 $ &$ 649 $& 3m6s & 254K & $<$1m & 469K\\
         \hline
         $\{1,4,6,8\}$ & $ 12 $ &$ 905 $& 4m58s & 499K & 4m & 2.6M\\
         \hline
         $\{1,5,8\}$ & $ 21 $ &$ 1533 $& 7m32s & 1.0M & 5h21m & 26M\\
         \hline
         $\{2,4,5,7\}$ & $ 13 $ &$ 594 $& 3m9s & 315K & 22m & 5.6M\\
         \hline
         $\{2,5,8\}$ & $ 27 $ &$ 1787 $& 8m36s & 1.3M & 22h15m &70M \\
         \hline
         $\{2,6,8\}$ & $ 31 $ &$ 3466 $& 9m2s & 2.5M & 24h32m &30M \\
         \hline
         $\{3,6,8\}$ & $ 31 $ &$ 3064 $& 8m34s & 2.3M & 17h17m &33M \\
         \hline
         $\{3,7,8\}$ & $ 27 $ &$ 2130 $& 7m54s & 1.4M & 24h19m &46M \\
         \hline
         $\{4,7,8\}$ & $ 21 $ &$ 2375 $& 7m37s & 1.4M & 15h35m & 28M\\
         \hline

    \end{tabular}
    \caption{The performance of all cuts in this pentagon-box example with numerator degree 5. The second and the third column is the number of master integrals and IBP identities, respectively. The fourth column is the {\NeatIBP} time used for generating the IBP system. The fifth column is the disk size of the IBP system generated by {\NeatIBP}. The sixth column is the time used for the {\Kira} reduction including the interface. The seventh column is the disk size of the reduced IBP table. Notice that, for the sixth column, we ran with ``\texttt{--parallel=20}''. Still, part of the 20 CPU threads were sometimes vacant in the reduction.}
    \label{tab: pentabox example spc deg5}
\end{table}

The merge step took 6 minutes. The merged result took $252$ MB of disk space. Still, we compared the spanning-cuts-merged result and the reduction result from the normal mode. They agree with each other.

\subsection{A four-loop non-planar massive example}
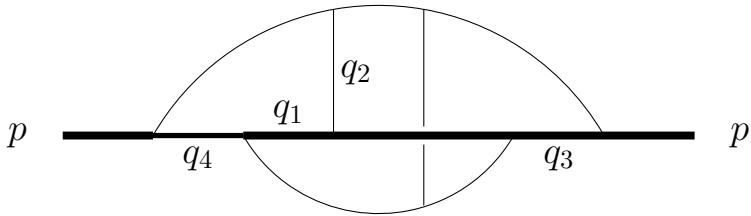
\begin{figure}[hbtp]
\centering
\begin{tikzpicture}[scale=0.6]
\draw[line width=3pt] (-2,0) -- (0,0);
\draw[line width=2pt] (0,0) -- (2,0);  
\draw[line width=3pt] (2,0) -- (12,0);
\draw (0,0) arc (150:30:5.77);
\draw (4,0)--(4,2.8);
\draw (2,0) arc (-150:-30:3.46);
\draw (6,-1.55) -- (6,-0.2);
\draw (6,2.8) -- (6,0.2);
\node[font=\large\bfseries] at (-3,0){$p$};
\node[font=\large\bfseries] at (13,0){$p$};
\node[font=\large\bfseries] at (3,0.5){$q_1$};
\node[font=\large\bfseries] at (4.5,1.4){$q_2$};
\node[font=\large\bfseries] at (9,-0.5){$q_3$};
\node[font=\large\bfseries] at (1,-0.5){$q_4$};
\end{tikzpicture}
\caption{A non-planar four-loop two-point diagram}
\label{fig:4-loop top decay}
\end{figure}

In this sub section, we choose a non-planar diagram to demonstrate the potential of the spanning cuts method in solving problems with high loop number. This example is related to the computation of N$^3$LO semileptonic b $\to$ u decay width and top quark decay width\footnote{See \cite{Fael:2023tcv,Chen:2023dsi} and reference therein for the recent progresses on these computations.}. The diagram is shown in Fig. \ref{fig:4-loop top decay}. 

The definition of propagators are
\begin{equation}
    \begin{aligned}
        &D_1=q_1^2-m_t^2,\quad
        D_2=q_2^2,\quad
        D_3=(q_1+q_2)^2-m_t^2,\quad
        D_4=q_3^2-m_t^2,\\
        &D_5=(q_1+q_2+q_3)^2,\quad
        D_6=(q_3-p)^2,\quad
        D_7=(q_2+q_4+p)^2,\\
        &D_8=(q_2 +q_3 + q_4)^2,\quad
        D_9=(q_1-q_4)^2,\quad
        D_{10}=q_1^2-m_W^2,\\
        &D_{11}=(q_4+p)^2,\quad
        D_{12}=(q_2+q_4)^2,\quad
        D_{13}=(q_3+q_4)^2,\\
        &D_{14}=(q_1+p)^2.
    \end{aligned}
\end{equation}
where $D_{12}\sim D_{14}$ are ISP. The kinematics condition is $p^2=m_t^2$. There are 6 target integrals from the top sector, with only $D_{10}$ the double propagator. The maximum numerator degree of the target integrals is 3. The input files can be found with example name ``2L4P\_massive''. 

We ran this example using the spanning cuts method in {\NeatIBP}. There are in total $29$ cuts. We picked up some of them as well as their performance, listed in Tab. \ref{tab: 4-loop example spc}.
\begin{table}[h]
    \centering
    \begin{tabular}{|c|c|c|c|c|}
    \hline
         cut & \#MI&\#IBP&running time&IBP disk size\\
         \hline
    
         $\{1,2,4,5\}$ & $84$ &$4217$& 8h7m& 4.8M \\
         \hline
         $\{1,2,4,7\}$ & $64$ &$3081$& 5h39m& 3.2M \\
         \hline
         $\{1,2,4,8\}$ & $77$ &$3761$& 5h21m& 3.2M \\
         \hline
         $\{1,2,5,7,8\}$ & $30 $ &$1313 $& 41m & 1.1M \\
         \hline
         $\{1,3,4\}$ & $152 $ &$14806 $& 12h44m & 17M \\
         \hline
         $\{1,4,5,7\}$ & $108 $ &$ 7313$& 10h57m & 13M \\
         \hline
         $\{2,3,5,6,7\}$ & $25 $ &$540 $& 42m & 292K \\
         \hline
         $\{2,5,6,9\}$ & $59 $ &$ 1401$& 4h9m  & 1.0M \\
         \hline
         $\{3,4,7,8\}$ & $61 $ &$4254 $&5h21m  & 4.2M \\
         \hline
         $\{3,5,6,7,9\}$ & $34 $ &$673 $& 1h0m & 426K \\
         \hline
         $\cdots$ & $\cdots$ &$\cdots$& $\cdots$ & $\cdots$ \\
         
    \end{tabular}
    \caption{The performance of some of the cuts in this four-loop example. The second and the third column is the number of master integrals and IBP identities, respectively. Since there are too many cuts, it is not neat to list all of them in this table. We have chosen the most easy one and the most hard one in this table, judged by their IBP disk size. The most easy one is cut $\{2,3,5,6,7\}$. The most complicated one is cut $\{1,3,4\}.$}
    \label{tab: 4-loop example spc}
\end{table}

This example is an example that consistency check does not pass. As stated in the previous sectors, this can happen for diagrams with so many massive propagators. Thus, there is no reduction step for this diagram. However, this example still indicates the advantage of the spanning cuts method. Without the spanning cuts method, this example is hard to be finished. It gets stuck even at the module intersection computation in {\Singular} in the top sector for several days, much longer than the total evaluation time using the spanning cuts method, which is approximately 13 hours. Even if the user uses the \texttt{flexible degree bound} mechanism introduced in Section \ref{sec: other feature}, it is still difficult. The {\Singular} indeed finished, at too-low degree bounds. The resulting incomplete \Gr basis leads to frequent usage of the module \texttt{AdditionalIBPs} in {\NeatIBP}. This makes the algorithm closer to Laporta, i.e. the denominator indices are severely increased. This makes it taking too much memory in the seeding step. 

In comparison with the problems without the spanning cuts method, we claim that this method at least provides us the potential of using the syzygy method in examples with such a high loop number. Once the cause for the inconsistency is found and the algorithm is adjusted in {\NeatIBP}, such kinds of examples will hopefully be solvable.

\subsection{A three-loop four-point massless example}
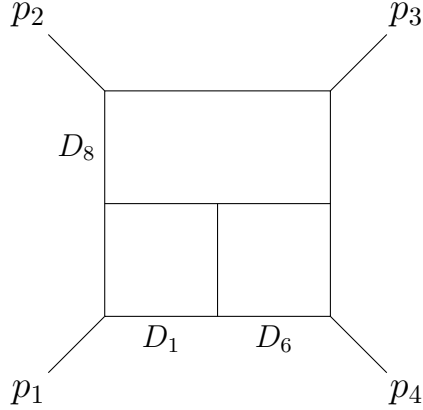
\begin{figure}[hbtp]
\centering
\begin{tikzpicture}[scale=0.75]
\draw (-1,-1) -- (0,0);
\draw (5,-1) -- (4,0);
\draw (-1,5) -- (0,4);
\draw (5,5) -- (4,4);
\draw (0,0) -- (4,0)--(4,4)--(0,4)--(0,0);
\draw (0,2) -- (4,2);
\draw (2,2) -- (2,0);

\node[font=\large\bfseries] at (-1.35,-1.35){$p_1$};
\node[font=\large\bfseries] at (-1.35,5.35){$p_2$};
\node[font=\large\bfseries] at (5.35,-1.35){$p_4$};
\node[font=\large\bfseries] at (5.35,5.35){$p_3$};
\node at (1,-0.4){{\color{black}$D_1$}};
\node at (3,-0.4){{\color{black}$D_6$}};
\node at (-0.5,3){{\color{black}$D_8$}};
\end{tikzpicture}
\caption{The massless three-loop four-point tennis court diagram}
\label{fig:tennis court}
\end{figure}

In this sub section, we exhibit a massless three-loop four-point example. This example can be found in the example folder stated above with the example name ``\texttt{tenniscourt}''. The diagrams is shown in Fig. \ref{fig:tennis court}.

The propagators are 
\begin{equation}
\begin{aligned}
    &D_{1}=l_1^2,\quad
    D_{2}=(l_1+p_1)^2,\quad
    D_{3}=(l_1-l_2)^2,\quad
    D_{4}=(l_1-l_3)^2,\\
    &D_{5}=(l_2-l_3)^2,\quad
    D_{6}=l_3^2,\quad
    D_{7}=(l_3+p_1+p_2+p_3)^2,\\
    &D_{8}=(l_2+p_1)^2,\quad
    D_{9}=(l_2+p_1+p_2)^2,\quad
    D_{10}=(l_2+p_1+p_2+p_3)^2,\\
    &D_{11}=l_2^2,\quad
    D_{12}=(l_1-p_2)^2,\quad
    D_{13}=(l_1-p_3)^2,\quad
    D_{14}=(l_3-p_1)^2,\\
    &D_{15}=(l_3-p_2)^2,
\end{aligned}   
\end{equation}
where $D_{11}\sim D_{15}$ are ISP. The kinematics conditions are
\begin{equation}
    \begin{aligned}
        &p_1^2=p_2^2=p_3^2=p_4^2=0,\\
        &(p_1+p_2)^2=s,\quad
        (p_1+p_4)^2=t,
    \end{aligned}
\end{equation}
where $p_1+p_2+p_3+p_4=0$.

The target integrals in this example are integrals in the top sector with numerator degree 5, without multiple propagators. In this example, we also ran in two modes, with and without spanning cuts. So, the symmetry relations are not included. 

We first ran this example without spanning cuts. {\NeatIBP} took about 13 hours and generated $112938$ IBP identities. The number of master integrals found is $88$. The disk size of the IBP identities is $72.7$ MB. After IBP generation, they are reduced using {\Kira} with {\FireFly}. The conversion of the IBPs to {\Kira} readable files took $3$ minutes. The reduction took 2 hours and 27 minutes. The disk size of the resulting reduced IBP table is $2.7$ MB. We did not use the parallelization in {\Kira} using ``\texttt{--parallel=...}'' in this example to provide a better estimation of the CPU resources needed for the reduction. We did the same in the spanning cuts mode while running this example.

Then, we ran this example with spanning cuts by adding 
\begin{lstlisting}[language=Mathematica]
SpanningCutsMode = True;
\end{lstlisting}
in the ``\texttt{config.txt}'' file. {\NeatIBP} found 22 cuts. Since there are too many of them, we selected some of the cuts and demonstrate their performance are shown in Tab. \ref{fig:tennis court}. The selection includes the most complicated cut $\{3,4,7,8\}$ and the most simple cut $\{1,3,5,6,8,10\}$, judged by the disk size of the IBP file generated by {\NeatIBP} and the IBP reduction time. In this table, the time used for spanning cuts consistency check and IBP shorten is not included. They took in total 3 minutes.

\begin{table}[h]
    \centering
    \begin{tabular}{|c|c|c|c|c|c|c|}
    \hline
         cut & \#MI&\#IBP&$T_{\text{gen}}$&IBP size&$T_{\text{red}}$ & reduced IBP size\\
         \hline
    
         $\{1,3,4,7,9\}$ & $ 21 $ &$ 10994 $& 47m & 7.1M & 13m & 133K\\
         \hline
         $\{1,3,5,6,8,10\}$ & $ 6 $ &$ 4146 $& 7m & 1.6M & $<$1m & 42K\\
         \hline
         $\{1,4,5,9\}$ & $ 23 $ &$ 14889 $& 39m & 8.4M & 14m & 130K\\
         \hline
         $\{2,3,4,7,10\}$ & $ 16 $ &$ 7203 $& 20m & 4.2M & 4m & 155K\\
         \hline
         $\{2,3,5,7\}$ & $ 22 $ &$ 10858 $& 26m & 6.0M & 5m & 86K\\
         \hline
         $\{2,4,5,10\}$ & $ 20 $ &$ 24952 $& 1h14m & 14M & 15m & 81K\\
         \hline
         $\{2,4,5,6,9\}$ & $ 21 $ &$ 4849 $& 33m & 3.1M & 3m & 300K\\
         \hline
         $\{2,4,5,7,8\}$ & $ 16 $ &$ 4425 $& 26m & 2.5M & 2m & 195K\\
         \hline
         $\{3,4,6,8,10\}$ & $ 9 $ &$ 14268 $& 14m & 6.1M & 4m & 42K\\
         \hline
         $\{3,4,7,8\}$ & $ 20 $ &$ 33014 $& 1h22m & 21M & 25m & 79K\\
         \hline
         $\cdots$ & $\cdots$ &$\cdots$& $\cdots$ & $\cdots$& $\cdots$& $\cdots$ \\

    \end{tabular}
    \caption{The performance of all cuts in this tennis-court example. The second and the third column is the number of master integrals and IBP identities, respectively. The fourth column is the {\NeatIBP} time used for generating the IBP system. The fifth column is the disk size of the IBP system generated by {\NeatIBP}. The sixth column is the time used for the {\Kira} reduction including the interface. The seventh column is the disk size of the reduced IBP table. Notice that, for the sixth column, we still did not turn on parallelization in {\Kira}, like what we did in the normal mode.}
    \label{tab: tenniscourt example spc}
\end{table}

After the reduction, {\NeatIBP} took less than a minute to merge the results. The merged IBP table takes 2.6 MB of disk space. The total time used, including generating, reducing and merging the IBP relations, was 1 hour and 51 minutes. The merged result agrees with the IBP table resulted in the normal mode (without spanning cuts).

Besides the above, we use this example to show the effect of the syzygy vector simplification algorithm introduced in Section \ref{sec: syzygy simplification algorithm}, running without spanning cuts. The additional settings added in ``\texttt{config.txt}'' is 
\begin{lstlisting}[language=Mathematica]
SimplifySyzygyVectorsByCut=True;
LiftSelectionSingularTimeConstraint=120
FurtherSelectionTimeUsedLimit=120
FurtherSelectionSingularTimeConstraint=15
\end{lstlisting}

We selected some of the sectors and show the difference of their performance before and after applying the simplification algorithms  in Tab. \ref{tab: comparing simplification in normal mode of 3L4P massless}

\begin{table}[h]
    \centering
    \begin{tabular}{|c|c|c|c|c|c|c|}
    \hline
          & \multicolumn{3}{|c|}{without syzygy simplification} & \multicolumn{3}{|c|}{with syzygy simplification}\\
         \hline
         sector& \#gen & time used& memory used& \#gen & time used& memory used \\
         \hline
          1023 & 666 & 1h8m & 20.2G & 16 & 15m & 6.7G \\
         \hline
         1022 & 665 & 4h4m & 31.2G & 39 & 1h4m & 18.8G \\
         \hline
          1021 & 536 & 1h21m & 22.0G & 25 & 13m & 4.0G \\
         \hline
          1019 & 541 & 50m & 14.6G & 30 & 9m & 2.9G \\
         \hline
          1015 & 654 & 1h17m & 16.9G & 21 & 16m & 5.0G \\
         \hline
          1013 & 602 & 2h1m & 17.1G & 32 & 37m & 11.3G \\
         \hline
          981 & 432 & 2h24m & 11.8G & 61 & 47m & 9.2G \\
         \hline
         949 & 520 & 2h1m & 13.6G & 514 & 3h37m & 20.1G \\
         \hline
         719 & 404 & 56m & 5.3G & 398 & 1h19m & 5.6G \\
         \hline
          511 & 721 & 38m & 10.8G & 31 & 13m & 4.4G \\
          \hline
          379 & 657 & 39m & 8.4G & 657 & 1h44m & 8.9G \\
         \hline
         351 & 778 & 2h13m & 13.4G & 778 & 2h33m & 14.0G \\
         \hline
         251 & 510 & 46m & 6.8G & 35 & 14m & 4.1G \\
         \hline
         $\cdots$ & $\cdots$ &$\cdots$& $\cdots$ & $\cdots$& $\cdots$& $\cdots$ \\
         
    \end{tabular}
    \caption{The comparison of the performance of the tennis-court diagram, before and after applying the simplification algorithm. The information includes the number of syzygy generator vectors (\#gen) before and after applying the simplification algorithm. It also includes the total time and memory used by {\NeatIBP} in those sectors. The sectors in the table are labeled by their binary code defined by $\sum_{i \in S}2^{i-1}$ for a sector $S$. For example, the top sector is labeled by $1023$.}
    \label{tab: comparing simplification in normal mode of 3L4P massless}
\end{table}

As shown in Tab. \ref{tab: comparing simplification in normal mode of 3L4P massless}, for many sectors, the numbers of syzygy generators were dramatically decreased. As a consequence, the time and memory spent on the corresponding evaluation also decreased considerably. Though, there are several sectors whose number of generators was nearly unchanged after applying the algorithms. This is because the computation time exceeds the user-set limit, so the program gave up corresponding simplification in these sectors. As a consequence, in these sectors, we did not benefit from the algorithms, but still wasted computation resources on running the simplification algorithms. Thus, the time and memory used increased in these sectors. We notice that, the time limit in the settings does not constrain the sorting step in the algorithms. Thus, the total time wasted in the simplification can still be large\footnote{Future versions of {\NeatIBP} may include some refinements on this point. }. Despite these unlucky sectors, the total time used for all the sectors still decreased after applying the simplification algorithms. It took 10 hours to finish. The number of IBP identities generated is $129881$. The disk size of the IBP identities is $81.0$ MB. The number of master integrals is $88$. Afterwards, the interface took 3 minutes to convert the IBP identities to {\Kira} readable files, and the reduction took 3 hours. The reduced IBP table takes $2.6$ MB of disk space. The result is checked and agrees with those after running without the simplification algorithms.

\section{Summary}\label{sec: summary}
In this paper, we present the new version (1.1) of {\NeatIBP}. One of the main new features in this version is the interface with {\Kira}. The interface allows the user to perform IBP reduction by the \textit{user-defined system} in {\Kira} using the small-sized IBP system generated by {\NeatIBP}, in a highly automated way. We also implemented the \textit{spanning cuts} method in this new version. This method, together with the interface with {\Kira}, forms a work flow to perform the IBP reduction on spanning cuts and merge them to a final result. We also introduced other new features in {\NeatIBP} 1.1 including the syzygy vector simplification algorithms. These new features provide {\NeatIBP} more potential to increase efficiency or to solve more difficult problems, including diagrams with higher loops or more parameters. 

We anticipate the following related work in the future.
\begin{enumerate}
    \item Solve the spanning cuts consistency problem. We will try to study the cause of the inconsistency and adjust the algorithms and codes of {\NeatIBP}, to make it fitting the diagrams that meets inconsistency.
    \item Interface more software in the market. We anticipate providing more choices for the IBP solver. 
    \item Currently, {\NeatIBP} relies on {\sc SpaSM} to solve linear systems on finite fields. Future development will try to provide more alternatives to the linear solver for users, including {\sc FiniteFlow} \cite{Peraro:2016wsq,Peraro:2019okx}.
    \item Develop a nice and efficient algorithm to cut multiple propagators.
    \item Try to adopt symmetry relations in the present of cut indices.
\end{enumerate}

\section*{Acknowledgement}
We thank Long-Bin Chen, Bo Feng, Fabian Lange, Roman Nikolaevich Lee, Yan-Qing Ma, Vasily Sotnikov, Hefeng Xu, Yongqun Xu, Simone Zoia for important discussions. ZW is supported by The First Batch of Hangzhou Postdoctoral Research Funding in 2024. The work of JB was supported by the Deutsche Forschungsgemeinschaft (DFG, German Research Foundation) - Project- ID 286237555 - TRR 195, Project B5, and Potentialbereich SymbTools - Symbolic Tools in Mathematics and their Application of the Forschungsinitative Rheinland-Pfalz. RM is supported by the Outstanding PhD Students Overseas Study Support Program of University of Science and Technology of China.  JU and YX are funded by the Deutsche Forschungsgemeinschaft (DFG, German Research Foundation) Projektnummer 417533893/GRK2575 “Rethinking Quantum Field Theory”. JU is funded by the European Union through the European Research Council under grant ERC Advanced Grant 101097219 (GraWFTy). Views and opinions expressed are however those of the authors only and do not necessarily reflect those of the
European Union or the European Research Council Executive Agency. Neither the European Union nor the granting authority can be held responsible for them. YZ is supported from the NSF of China through Grant No. 12247103, and thanks the Galileo Galilei Institute for Theoretical Physics for the hospitality and the INFN for partial support during the completion of this work. ZW expresses gratitude to the other authors for their agreement on the ordering of the author names.

\section*{Declaration of generative AI and AI-assisted technologies in the writing process} 
During the preparation of this work, the authors used Writefull Premium to perform spelling/grammar checks, and to improve the language and readability of this paper. After using this service, the authors reviewed and edited the content as needed and take full responsibility for the content of the publication.

\bibliographystyle{elsarticle-num}
\bibliography{bibtex}

\end{document}